\documentclass[fleqn,usenatbib]{mnras}

\usepackage[T1]{fontenc}
\usepackage{ae,aecompl}

\usepackage{graphicx}	
\usepackage{amsmath}	
\usepackage{amssymb}	

\title[The nature of the first BH seeds]{Light, medium-weight or heavy? The nature of the first supermassive black hole seeds}
\usepackage{newtxtext,newtxmath}
\author[Sassano et al.]{Federica Sassano$^{1,2}$\thanks{Contact e-mail: \href{mailto:federica.sassano@uniroma1.it}{federica.sassano@uniroma1.it}}, Raffaella Schneider$^{1,2,3}$, Rosa Valiante$^{2,3}$, Kohei Inayoshi$^{4}$,
\newauthor
Sunmyon Chon$^{5}$, Kazuyuki Omukai$^{5}$, Lucio Mayer$^{6}$ and Pedro R. Capelo$^{6}$
\\
$^{1}$Dipartimento di Fisica, Sapienza, Universit$\grave{a}$ di Roma, Piazzale Aldo Moro 5, 00185, Roma, Italy  \\
$^{2}$INFN, Sezione di Roma I, P.le Aldo Moro 2, 00185 Roma, Italy  \\
$^{3}$INAF/Osservatorio Astronomico di Roma, Via di Frascati 33, 00078 Monte Porzio Catone, Italy\\
$^{4}$Kavli Institute for Astronomy and Astrophysics, Peking University, Beijing 100871, China\\
$^{5}$Astronomical Institute, Graduate School of Science, Tohoku University, Aoba, Sendai 980-8578, Japan \\
$^{6}$ Center for Theoretical Astrophysics and Cosmology, Institute for Computational Science, University of Zurich, Winterthurerstrasse\\ 190,  CH-8057, Z{\"u}rich, Switzerland}

\date{Accepted 2021 June 14. Received 2021 June 14; in original form 2020 August 3}

\pubyear{2020}

\begin{document}
\label{firstpage}
\pagerange{\pageref{firstpage}--\pageref{lastpage}}
\maketitle

\begin{abstract}
Observations of hyper-luminous quasars at $z>6$ reveal the rapid growth of supermassive black holes (SMBHs $>10^9 \rm M_{\odot}$) whose origin is still difficult to explain. 
Their progenitors may have formed as remnants of massive, metal free stars (light seeds), via stellar collisions (medium-weight seeds) and/or massive gas clouds direct collapse (heavy seeds).
In this work we investigate for the first time the relative role of these three seed populations in the formation of $z>6$ SMBHs within an Eddington-limited gas accretion scenario. To this aim, we implement 
in our semi-analytical data-constrained model a statistical description of the spatial fluctuations of Lyman-Werner (LW) photo-dissociating radiation and of metal/dust enrichment. This allows us to set the physical conditions for BH seeds formation, exploring their relative birth rate in a highly biased region of the Universe at $z>6$.
We find that the inclusion of medium-weight seeds does not qualitatively change the growth history of the first SMBHs: although less massive seeds ($<10^3 \rm M_\odot$) form at a higher rate, the mass growth of a $\sim 10^9 \rm M_\odot$ SMBH at $z<15$ is driven by efficient gas accretion (at a sub-Eddington rate) onto its heavy progenitors ($10^5 \rm M_\odot$). This conclusion holds independently of the critical level of LW radiation and even when medium-weight seeds are allowed to form in higher metallicity galaxies, via the so-called super-competitive accretion scenario. Our  study  suggests  that  the  genealogy  of $z \sim 6$ SMBHs  is  characterized  by  a  rich  variety  of  BH  progenitors, which represent only a small fraction ($< 10 - 20\%$) of all the BHs that seed galaxies at $z > 15$.
\end{abstract}




\section{Introduction}
\label{section:introduction}

Starting from their first detection by \citet{fan2002, fan2003}, more than 100 supermassive black holes (SMBHs), with masses $\geq 10^9 \rm  M_\odot$, have been observed at $z > 6$ as bright quasars 
\citep[e.g.][]{willott2010, Jiang2016, Banados2016, matsuoka2019discovery, yang2020}.
The most massive among this high-redshift sample is the hyper-luminous quasar SDSS J0100+2802, 
with a mass of $1.2 \times 10^{10} \rm M_\odot$ at $z = 6.3$ \citep{wu2015}. The three most distant ones are J0313-1806 \citep{wang2021luminous}, with a SMBH mass of $1.6 \times 10^9 \rm M_{\odot}$,   ULAS J1342+0928 \citep[][]{banados2018800}, with a SMBH mass of $7.8 \times 10^8 \rm M_{\odot}$ detected at $z=7.54$, and the recently reported {\it P$\bar{o}$niu$\bar{a}$'ena} (J1007+2115, \citealt{yang2020}), with a SMBH mass of
$1.5 \times 10^9 \rm  M_\odot$ at $z = 7.52$, when the Universe was only $0.69$ Gyr old. The estimated SMBH masses imply that common mechanisms of Eddington-limited gas accretion onto stellar-mass black holes fail to provide a fast enough BH mass growth at these early cosmic epochs (see, for a recent review, \citealt{inayoshi2020}).

A possible scenario is to consider Population III (Pop III) stellar remnants, or \textit{light seeds}, as progenitors of SMBHs. Pop III stars are expected to form around $z\sim 20-30$ \citep[][]{bromm2013}, in the first collapsed structures, \textit{minihalos} with virial temperatures $T_{\rm vir}<10^4$K where the collapse of primordial gas is driven by molecular hydrogen cooling. While hydrodynamical models assuming  spherical symmetry \citep[][]{omukai1998formation,omukai2010low} and those starting from cosmological initial conditions \citep[][]{yoshida2008} show a remarkable agreement in describing the early collapse phase and the formation of the central hydrostatic core (see also \citealt{bromm2013} and references therein), large uncertainties still remain on the evolution of the post-collapse phase, 
when disk fragmentation, protostellar evolution, and radiative feedback become important 
(\citealt{omukai2003formation,hosokawa2011protostellar}; see for a review \citealt{greif2015}). The resulting Pop III mass distribution is generally found to be top-heavy, with stellar masses ranging from a few tens to several hundreds of solar masses
\citep[][]{hirano2014one, hirano2015, hosokawa2016} and a non-negligible fraction of stars formed in binary or multiple systems
\citep[][]{sugimura2020}. Independent constraints on the Pop III initial mass function (IMF) come from stellar archaeology studies
(see \citealt{frebel2015} for a thorough review). The low-metallicity tail of the metallicity distribution function of Galactic halo stars and their observed large carbon-to-iron surface abundance ratios appear to be consistent with models where 
Pop III stars form in the range $[10 - 300]\, \rm M_\odot$ with a characteristic mass of 20 $\rm M_\odot$ \citep[][]{de2016limits} and where the early chemical enrichment is dominated by the explosion of faint supernovae (SNe, \citealt{marassi2014}). 
Even assuming that BH remnants of this first stellar generation can extend to masses of a few 100s of $\rm M_{\odot}$, these light seeds
can grow in mass to form SMBHs at $z > 6$ only if they accrete gas at super-Eddington rates, even in short and intermittent phases
\citep[][]{pezzulli2016,pezzulli2017sustainable}. Recently, simulations have become capable of studying super-critical accretion in realistic astrophysical environments, assuming  slim accretion disks \citep[][]{Sadowski2009, madau2014} or spherically accreting envelopes, without dynamically important angular momentum  \citep[][]{inayoshi2016,2020MNRAS.tmp.2035T}.
The question is whether the high gas densities required to sustain Super-Eddington accretion can be found in light seed environments at high redshifts (see for a critical discussion \citealt{mayer2019}).

An alternative scenario to grow SMBHs in the early Universe is to start from \textit{heavy seeds}, $\sim 10^5 \rm M_{\odot}$ BHs formed
from the collapse of supermassive stars (SMSs) \citep[][]{omukai2001, bromm2003formation, wise2008, regan2009, shang2010supermassive, hosokawa2012, latif2013, 2014MNRAS.445L.109I, regan2014, becerra2015, chon2016, latif2016,   becerra2018,  wise2019,  maio2019}  or in gas-rich galaxy mergers \citep[][]{mayer2015, mayer2017, mayer2019}. In the first case, gas cooling and fragmentation must be avoided to form a single supermassive object. For this reason, \textit{heavy seeds} are expected to form in pristine halos with virial temperatures $T_{\rm vir}\geq 10^4 K$, the so-called \textit{atomic cooling halos}, where fragmentation is avoided because of the lack of metals and molecular hydrogen cooling is suppressed by a sufficiently strong radiation field in the Lyman-Werner (LW) band \citep[][]{omukai2008can, dijkstra2014feedback,sugimura2014critical, wolcottgreen2017, wolcottgreen2020suppression} or by dynamical heating caused by rapidly growing dark matter halos \citep[][]{mayer2015, wise2019}. In addition to avoiding fragmentation, very high accretion rates, with $\dot{M} > 0.01 \rm M_{\odot} yr^{-1}$, are required to reduce radiative feedback from the growing protostar onto the accretion flow \citep[][]{hosokawa2012, chon2018,matsukoba2019gravitational}. 

Finally, a third scenario foresees the formation of BH seeds with mass $\sim 10^3 \rm M_\odot$, also called Intermediate Mass Black Holes (IMBHs), that we will refer to as \textit{medium-weight seeds}. Their formation is expected to occur through run-away stellar collisions in dense star clusters \citep[][]{omukai2008can, volonteri2010formation, davies2011supermassive, devecchi2012, katz2015, sakurai2017formation, stone2017formation, reinoso2018, tagawa2020making}. These are supposed to originate in atomic cooling halos where the gas initially undergoes an
almost isothermal collapse, similar to the early formation phase of \textit{heavy seeds}. Hence, their formation requires conditions similar to the ones
discussed above, lack of metal fine structure line and H$_2$ cooling. However, a dense star cluster may form provided that dust cooling drives fragmentation in the late phase of the collapse \citep[][]{omukai2008can}. It is important to stress that this simple picture may be significantly
affected by the dynamics of the infalling gas. Indeed, \citet{chon2020supermassive} show that, in the absence of H$_2$ cooling, although dust cooling promotes
the formation of a few thousand low-mass stars, the strong accreting flow preferentially feeds the central star which grows supermassive.  Hence, in this
\textit{super competitive accretion} (SCA) mode, the formation of \textit{heavy seeds} may continue even in moderately enriched halos, and 
\textit{medium-weight seeds} form only when metal line cooling starts to decrease the gas temperature, at $Z \sim 10^{-3} Z_\odot$, lowering the accretion rate.

The importance of \textit{medium-weight seeds} is not only related to SMBH genealogy, but has a number of additional implications in astrophysics. Medium-weight seeds would extend the correlation between the stellar and nuclear black hole masses to dwarf galaxies \citep[][]{greene2010, safonova2010extrapolating, reines2015, mezcua2018}, they could explain the existence of Ultra Luminous X-ray sources \citep[][]{miller2004comparison,fritze2018searching,shen2019fast, barrows2019,baldassare2020populating}, and are necessary to prevent tidal distruption events in young star clusters observed in the galactic nucleus
\citep[][]{kim2004dynamical}. 
These $\sim 10^3 \rm M_\odot$ BHs will be targeted by future third-generation gravitational telescopes, such as the Laser Interferometer Space Antenna (LISA) \citep[][]{coleman2002production,sesana2005gravitational,amaro2010detection,kremer2019post}, but recent observations \citep[][]{ballone2018weighing,argo2018searching,takekawa2019indication, nguyen2019, woo2019,barack2019black} have found evidence of their existence in the Local Universe.

In this work, we aim at assessing the relative importance of \textit{light, medium-weight} and \textit{heavy seeds}, in the mass growth histories of $z \sim 6$ SMBHs. To this aim, we have largely improved our original seeding prescription described in \citet{valiante2016first} and \citet{valiante2018statistics} in order to investigate the formation sites of all these three families of BH seeds in the cosmological evolution of a typical $z \sim 6$ SMBH and its host galaxy.
In particular, we present here a more physical treatment of the inhomogeneous radiative and chemical feedback.
We quantify the statistical distributions of all BH seeds in several independent simulations of the final SMBH, and explore how these are impacted by assuming different conditions (critical LW flux, metallicity, and dust-to-gas mass ratios) for their formation.

The paper is organized as follows. In section \ref{section:model} we briefly describe the semi-analytical model adopted in our analysis. 
Section~\ref{section:newFeatures} presents the new features implemented for this work, i.e. \textit{the medium-weight seed} formation scenario (sections~\ref{mwSeeds} and \ref{subsection:scamodel}) and the inhomogeneous treatment of metal enrichment and radiative LW flux (section \ref{section:chemevo}). In section \ref{section:results}, we describe the main results, that will be discussed and summarized in sections \ref{section:discussion} and \ref{section:summary}. 

For our analysis, we will assume a Planck Cosmology with $\Omega_m=0.314$, $\Omega_{\Lambda}=0.686$, $n_s=0.96$, $h=0.674$ and $\sigma_8=0.834$ from \citet{ade2014planck}.

\section{Description of the model}
\label{section:model}

Here we summarize the main features of the semi-analytical, data constrained model, \texttt{GAMETE/QSOdust} (\texttt{GQd}) developed to simulate the cosmological build-up of  a $z \sim 6$ SMBH and its host galaxy \citep[][]{valiante2011origin,valiante2014high,valiante2016first}.

\texttt{GQd} follows the co-evolution of nuclear BHs and their host galaxies and, at the same time, the metal and dust enrichment of their interstellar medium (ISM), in a cosmological context. For this reason, \texttt{GQd} is suitable to investigate the birth environments of BH seeds in high-redshift galaxies, their subsequent mass growth through gas accretion and mergers, and their
relative contribution to the final SMBH mass at $z \sim 6$. 
We refer the reader to
\citet{valiante2016first} (and references therein) for a detailed description of the original model.

\subsection{Dark matter halos evolution}
For a statistical analysis we reconstruct ten merger histories of a dark matter (DM) halo of $ M_{\rm halo}=10^{13}\rm M_{\odot}$ at $z_0=6.42$, presumably hosting a SMBH \citep[][]{ willott20033x10,cubbon2007temporal}.
We perform a binary Monte Carlo algorithm to decompose, backward in time, the massive DM halo into its progressively less massive progenitors, from $z_0$ back to $z=24$.

At a given redshift $z$, the resolution mass of the simulated merger trees, i.e. the minimum mass of a virialized structure, is described as in \citet{Valiante20}:

\begin{equation}
    M_{\rm res}(z)=10^{-3}M_{\rm halo}\biggl(\frac{1+z}{1+z_0}\biggr)^\beta, 
    \numberwithin{equation}{section}
\end{equation}\label{eq:resolutionMass}

\noindent
where $\beta=-7.5$ \citep[][]{valiante2016first}, so that $M_{\rm res}\sim 10^{6} \,\, (10^{10}) \, \rm M_\odot$ at $z=24$ ($z=z_0$).
Lower mass fragments, i.e. unresolved structures with $M<M_{\rm res}$, account for the external medium from which halos accrete mass, that we refer to as intergalactic medium (IGM). The functional form in Eq.~\ref{eq:resolutionMass} and the redshift-dependent characteristic time interval, $\Delta t$ (ranging between $0.2$ and 2 Myr), are designed to ensure the binarity of the merger tree ($<2$ progenitors per halo), to
resolve the formation of high-$z$ minihalos, to reproduce the halo mass function predicted by the Extended Press-Schechter (EPS) formalism \citep{lacey1993merger} and to contain computational costs.

\subsection{Star formation and feedback}
\label{section:SFR}

\subsubsection{Star formation law}
Along a merger tree, each progenitor galaxy can form stars according to the available gas budget, $M_{\rm gas}$. The star formation rate (SFR) is described as:
\begin{equation}
SFR = f_{\rm cool} \, M_{\rm gas}\, \epsilon/\tau_{\rm dyn}(z),
\label{eq:sfr}
\end{equation}
\noindent
where $\tau_{\rm dyn}(z)$ is the halo dynamical time, $\epsilon = \epsilon_{\rm quiesc}+\epsilon_{\rm burst}$ is the sum of the quiescent and merger-driven starburst efficiencies and is one of the model free parameters, that have been set to reproduce the observed properties of a proto-typical $z \sim 6$ QSO and its host galaxy (see Table \ref{tab:table1} for the complete set of free  parameters). We assume stars to form with a constant efficiency $\epsilon_{\rm quiesc} = 0.1$ that can be enhanced to $\epsilon_{\rm quiesc} + \epsilon_{\rm burst}$ during galaxy major mergers due to a reduction in the timescales for star formation from $\tau_{\rm quies}=\tau_{\rm dyn}/\epsilon_{\rm quiesc}$ to $\tau_{\rm burst}=\tau_{\rm dyn}/(\epsilon_{\rm burst}+\epsilon_{\rm quiesc})$. Following \citet{valiante2011origin}, we assume $\epsilon_{\rm burst}$ to be a function of the ratio $\mu$ between the mass of the less massive halo over the more massive companion and to be described by a normalized Gaussian distribution:
\begin{equation}
 \epsilon_{\rm burst}=\frac{e^{-(\mu-\mu_{\rm crit})^2/2\sigma^2_{\rm burst}}}{\sqrt{2\pi}\sigma_{\rm burst}}
\end{equation}
\noindent
with $\sigma_{\rm burst}=0.05$, $\mu_{\rm crit}=1$, and $\mu > 1/4$, which defines the threshold for galaxy major mergers. For equal mass mergers ($\mu=\mu_{\rm crit}$), this leads to a maximum value of $\epsilon_{\rm burst}=8$, reducing the timescale for star formation, $\tau_{\rm burst}/\tau_{\rm quies}= \epsilon_{\rm quiesc}/(\epsilon_{\rm quiesc} + \epsilon_{\rm burst}) \sim 0.012$.

Finally, the parameter $f_{\rm cool}$ in Eq. \, \ref{eq:sfr} quantifies the reduced cooling efficiency of minihalos with respect to atomic cooling halos, that we discuss below.

\subsubsection{Radiative feedback}
In each galaxy, the efficiency of gas cooling and star formation is regulated by radiative feedback. In our model we account for the effects of both photo-dissociating and photo-heating feedback. 

\begin{figure}
     \centering
     \includegraphics [scale=0.34]{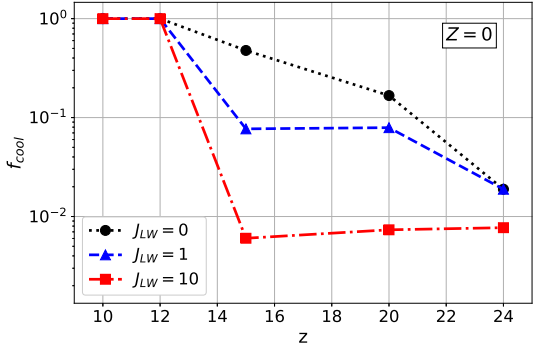}
     \caption{The predicted reduced cooling efficiency $f_{\rm cool}$ in minihalos as a function of redshift. The black, blue, and red dashed lines correspond to different values of the LW flux ($J_{\rm LW} = 0, 1, 10$, respectively) illuminating halos with mass equal to the resolution mass of the simulation, $M_{\rm vir} = M_{\rm res}(z)$, and with a constant gas metallicity $Z = 0$. The resulting cooling efficiency is independent of metallicity for poorly enriched ($Z\leq 10^{-2} \, Z_\odot$) minihalos.}
     \label{fig:fcool1}
 \end{figure}

 At low metallicity, the gas cooling efficiency in minihalos relies only on molecular hydrogen, that can be easily dissociated by photons in the LW band. 
We account for this effect through the parameter $f_{\rm cool}$ entering in Eq. \, \ref{eq:sfr}.
In particular, we set $f_{\rm cool}=1$ in atomic cooling halos, whereas in minihalos its value ($\leq 1$) depends on the halo virial temperature, $T_{\rm vir}$, redshift, gas metallicity, and intensity of the illuminating 
LW flux, $J_{\rm LW}$, expressed in units of $10^{-21}$ erg/s/Hz/cm$^2$/sr \citep[see][for details]{valiante2016first, de2016limits}. 
 In Fig.\, \ref{fig:fcool1} we show the redshift evolution of the reduced cooling efficiency for the least massive halos (for halos with $M_{\rm vir} = M_{\rm res}(z)$) assuming 
three different values of the LW irradiating flux, $J_{\rm LW} = 0, 1,$ and $10$, and a constant gas metallicity equal to $Z = 0$. Similar results are found for gas metallicity in the range $0\leq Z \leq 10^{-2} \, Z_\odot$.  The figure shows that even when $J_{\rm LW} = 0$, $f_{\rm cool}$ ranges between 0.02 to 1 due to less efficient cooling by molecular hydrogen. At $z > 12$, $M_{\rm res}(z)$ corresponds to mini-halos and $f_{\rm cool}$ is very sensitive to the illuminating LW radiation field. At $z \le 12$, $M_{\rm res}(z)$ grows above the halo mass with $T_{\rm vir} = 10^4$ K and $f_{\rm cool} = 1$ in atomic cooling halos. When $J_{\rm LW} = 100$, $f_{\rm cool}=0$ at all redshifts greater than $12$ and star formation in mini-halos is completely suppressed.

 Even in atomic cooling halos star formation can be inhibited as a consequence of the increased gas temperature within photo-ionized regions. We assume $\epsilon = 0$ in Eq. \ref{eq:sfr} when the halo virial temperature is below the IGM temperature, $T_{\rm vir} < T_{\rm IGM}$. The latter is computed as:
\begin{equation}
    T_{\rm IGM} = Q_{\rm HII}(z) T_{\rm reio} + [1-Q_{\rm HII}(z)] T_{\rm gas}
    \label{eq:IGMtemp}
\end{equation}
\noindent
where $Q_{\rm HII}(z)$ is the filling factor of HII regions, $T_{\rm reio} = 2\times 10^4$ K is the post-reionization temperature and $T_{\rm gas}=170$ K $[(1+z)/100]^2$.
We compute the time evolution of $Q_{\rm HII}(z)$ as:
\begin{equation}
    \frac{d Q_{\rm HII}}{dt} = f_{\rm esc} \dot n_{\gamma}/n_{\rm H} - \alpha_B \, C \, n_{\rm H} (1+z)^3 Q_{\rm HII}
\end{equation}
\noindent
where $f_{\rm esc}=0.1$ is the escape fraction of ionizing photons, $\dot n_{\gamma}$ is the total production rate of ionizing photons per unit volume summed over all the emitting sources, $n_{\rm H}$ is the comoving hydrogen number density in the IGM, $\alpha_B = 2.6 \times 10^{-13}$ cm$^3$ s$^{-1}$ is the hydrogen recombination rate and
$C = 3$ is the clumping factor (for more details, see \citealt{valiante2016first}).

\subsubsection{Chemical and mechanical feedback}

In \texttt{GQd} stars form according to a Larson IMF \citep[][]{larson1998early}:
\begin{equation}
    \phi(m_*)\propto m_*^{(\alpha-1)}e^{-m_{\rm ch}/m_*},
\end{equation}
\noindent
where $m_*$ is the stellar mass, $m_{\rm ch}$ is
the characteristic mass and $\alpha = -1.35$.
In metal free/poor ($Z < Z_{\rm cr}=10^{-3.8} \, Z_{\odot}$) environments we adopt an IMF constrained by stellar archaeology data \citep[][]{de2014decoding}, with Pop III stars forming in the mass range $[10 - 300] \, \rm M_{\odot}$ and $m_{\rm ch} = 20 \, \rm M_{\odot}$.
In addition, when the star formation efficiency is low, we randomly sample the above IMF to predict the stellar mass distribution present in each system \citep[see][for details]{valiante2016first}. Once the metallicity increases above the threshold $Z > Z_{\rm cr} = 10^{-3.8}Z_{\odot}$, metal fine-structure line cooling becomes efficient \citep{omukai2005thermal} and we assume Pop II/I stars to form in the range $[0.1 - 100] \, \rm M_\odot$ with $m_{\rm ch} = 0.35 \, \rm M_{\odot}$.

In each galaxy, evolving stars progressively enrich the ISM with metals and dust. Their abundances are computed adopting the grids of stellar yields provided by \citet[][]{vdHG97} and \citet[][]{ZGT08} for AGB stars ($1-8$ M$_\odot$), \citet{WoosleyWeaver95} and \citet[][]{BianchiSchneider09} for core-collapse SNe ($10-40$ M$_\odot$) and \citet{HegerWoosley02} and \citet[][]{BianchiSchneider09} for pair-instability SNe (PISNe, $140-260$ M$_\odot$).  For our adopted Pop III IMF, the largest contribution to the metal and dust yields is provided by the most massive stars (see the left panel of Fig. 2 in \citealt{de2014decoding}), for which the assumption of instantaneous recycling is a good approximation. Hence, Pop III stars are assumed to evolve instantaneously, i.e. their lifetime is the characteristic time-interval of the merger tree, while Pop~II/I stellar lifetimes are computed according to the parametric form proposed by \citet{raiteri1996study} and depend on the stellar mass and metallicity. Once injected in the ISM, the abundance of dust grains is computed accounting for grain destruction by SN shocks and grain growth in the dense, cold ISM phase \citep[for additional details, see][]{valiante2014high}.

\begin{table}
\caption{Free parameters of the Reference and SCA models. 
$\epsilon_{\rm quiesc}$ is the star formation efficiency (see Eq. \ref{eq:sfr}), $\epsilon_{\rm w,SN}$ is the SN-driven wind efficiency (see Eq. \ref{eq:mejsn}), $\alpha_{\rm BH}$
quantifies the efficiency of BH accretion (see Eq. \ref{eq:BHL}) and $\epsilon_{\rm w,AGN}$ is the BH-driven wind efficiency (see Eq.\ref{eq:dmgrb}). For each set of models, these values have been calibrated to reproduce the observed properties of the $z = 6.42$ QSO J1148 and its host galaxy (see \citealt{valiante2016first}).} 
\begin{tabular}{ccccc}
\hline
  Model & $\epsilon_{\rm quiesc}$ & $\epsilon_{\rm w,SN}$  & $\alpha_{\rm BH}$& $\epsilon_{\rm w,AGN}$ \\
  \hline
  Reference & $0.1$ & $1.6\cdot 10^{-3}$ & $150$ & $2.5\cdot 10^{-3}$  \\
  SCA & $0.1$ & $1.6\cdot 10^{-3}$ & $80$ & $2.5\cdot 10^{-3}$  \\
\hline
\end{tabular}
\label{tab:table1}
\end{table}

Finally, SN explosions deposit energy in the ISM, driving a galaxy-scale wind (mechanical feedback), 
\begin{equation}
\frac{dM_{\rm ej,SN}(t)}{dt}= \frac{2 E_{\rm SN} \epsilon_{\rm w,SN} R_{\rm SN} (t)}{v_{\rm e}^2},
\label{eq:mejsn}
\end{equation}
\noindent
that is proportional to the rate of SN explosions, $R_{\rm SN} (t)$, to the average SN explosion energy,
$E_{\rm SN}$, and to an efficiency parameter, $\epsilon_{\rm w,SN}$, that accounts for the fact that only a small fraction of the total energy injected by the SNe is in kinetic form and coupled to the gas, while the rest is dissipated and converted into thermal energy \citep[see e.g.][]{walch2015}. The quantity $R_{\rm SN} (t)$ depends on the SFR and on the stellar IMF, $E_{\rm SN} = 2.7 \cdot 10^{52}$\, erg for Pop III stars and
$1.2 \cdot 10^{51}$ erg for Pop II/I stars, and $\epsilon_{\rm w,SN}$ 
is a free parameter that we calibrate based on the observations (see Table \ref{tab:table1}).

\subsection{BH seeding prescriptions}
\label{section:BHseeds}
As introduced in section \ref{section:introduction}, seeds expected to form under specific environmental conditions (level of illuminating LW radiation, metal and dust composition of the ISM). In \texttt{GQd} such properties are computed consistently with the overall evolution of the galaxy and nuclear black hole populations \citep[][]{valiante2016first, valiante2018statistics, valiante2018observability}.

\noindent \textit{Light seeds} form as end-products of Pop III stars, in minihalos or atomic cooling halos where gas cooling is dominated by H$_2$. This implies that the illuminating LW flux must be sub-critical, $J_{\rm LW} < J_{\rm cr}$, and that metal- and dust-cooling must not operate, hence
the metallicity $Z < Z_{\rm cr}$ and the gas-to-dust mass ratio $\mathcal{D} < \mathcal{D_{\rm cr}}$. 
In these conditions, the star formation efficiency is relatively low and only a small number of stars is formed. The mass of the BH remnants depends on the initial mass of the (stochastically sampled) stars \citep[][]{woosley2002evolution}:
stars in the mass ranges $[40 - 140]\, \rm M_{\odot}$ and $[260 - 300] \rm M_\odot$ directly collapse into a BH with a mass equal to the progenitor mass; in the mass range $[140 - 260]\, \rm M_{\odot}$, the stars explode as Pair Instability SN (PISN) and leave no remnant. Depending on the star formation efficiency, these different mass ranges are unequally sampled and therefore the emerging BH mass distribution is not 
unique. 

We assume that the heaviest among all the formed BHs migrate to the centre by dynamical friction and we tag this as the nuclear \textit{light seed}. 
If a halo hosts multiple Pop III star formation episodes, we repeat the random sampling procedure and we assume the heaviest
among the new BH population to migrate to the centre and to merge with the previous nuclear BH, generating a new heavier \textit{light seed}.
\newline
\newline
\noindent \textit{Heavy seeds} mediated by the collapse of a SMS require the gas to monolithically collapse with no fragmentation \citep[][]{omukai2008can}. This, in turn, requires the gas to be hosted in atomic cooling halos that are illuminated by a sufficiently strong LW flux to inhibit H$_2$ cooling and, similarly to \textit{light seeds}, metal and dust cooling must not operate.
In this scenario we plant a \textit{heavy seed} of $10^5 \, \rm M_\odot$ in atomic cooling halos where $J \ge J_{\rm cr}$, and $Z < Z_{\rm cr}$ and $\mathcal{D} < \mathcal{D_{\rm cr}}$. In addition, we require the halo not to have experienced previous episodes of star formation and to have a sufficiently large reservoir of gas $M_{\rm gas} \geq 10^6 \,  \rm M_{\odot}$ (condition required to fuel strong accretion onto the forming SMS).
\newline
\newline
\noindent Following \citet[][]{valiante2016first}, in our reference model (that we call R300) we assume $J_{\rm cr} = 300$ (expressed in units of $10^{-21}$ erg/s/Hz/cm$^2$/sr), $Z_{\rm cr} = 10^{-3.8}Z_{\odot}$,
and $\mathcal{D_{\rm cr}} = 4.4\cdot 10^{-9}$ \citep[][]{schneider2012}.

\subsection{Black hole growth and feedback}
\label{section:BHgrowth}
The model \texttt{GQd} follows the mass growth of nuclear black holes through gas accretion and the coalescence with other black holes, starting from a seed population that depends on the environmental conditions as discussed above.

We assume that two nuclear BHs coalesce only in major merger events, i.e. when $\mu \ge 1/4$ \footnote{We assume a galaxy merger to occur within the redshift-dependent timestep of the simulation, $\Delta t$, which ranges from a fraction to a few Myr.}.
Although post-merger BH ejection due to asymmetric GW emission may have an impact on the SMBH growth, especially at high z \citep[see e.g.][]{pezzulli2016}, in the present analysis we neglect the gravitational recoil effect in other to compare our results with \citet{valiante2016first} but focusing on the relative role of all three BH seeds populations for $z > 6$ SMBHs growth.

In minor halo-halo mergers ($\mu < 1/4$) we instead assume that only the most massive BH will settle in the nuclear region of the newly formed galaxy, where it can start accreting gas. The lighter one is considered as a satellite, wandering in the galaxy outskirts, and we do not follow its subsequent evolution.
Such a conservative assumption is independent of the mass ratio of the two BHs involved in the galaxy merger event. However, during wet (gas-rich) minor galaxy mergers (down to 1:10), the satellite BH may grow faster than the primary so that it may orbit towards the centre as well \citep[see e.g.][]{Callegari2009, Callegari2011, van2014nuclear, Capelo2015}. We plan to investigate this effect, together with a more refined description of BH dynamics (including gravitational recoil) in future works.

Thus, not all the nuclear BHs present at $z> 6.4$ will directly contribute to the final SMBH mass, as some of these may be involved in minor halo-halo mergers and become wandering BHs at later times. The real SMBH progenitors are identified by reconstructing the assembly history backwards in time up to the seeds formation epoch.

As in \citet{valiante2011origin, valiante2014high, valiante2016understanding}, BH mass growth by gas accretion proceeds at a rate given by $\dot M_{\rm acc}=$min$[\dot M_{\rm Edd},\dot M_{\rm BHL}]$, where 
$\dot M _{\rm Edd}$ is the Eddington rate, 

\begin{equation}
\dot M_{\rm Edd} = \frac{4 \pi G M_{\rm BH} m_{\rm p}}{\epsilon_{\rm r} \sigma_{\rm T} c},
\end{equation}
\noindent
and $\dot M_{\rm BHL}$ is the Bondi--Hoyle--Lyttleton accretion rate \citep[][]{hoyle39,bondi1944mechanism,bondi1952spherically},

\begin{equation}
\dot M_{\rm BHL} = \frac{4 \pi \alpha_{\rm BH} G^2  M^2_{\rm BH} \rho_{\rm gas}(r_{\rm A})}{c^3_{\rm s}}.
\label{eq:BHL}
\end{equation}
\noindent
In the above equations, $m_{\rm p}$ is the proton mass, $\epsilon_{\rm r} = 0.1$ is the radiative efficiency, $\sigma_{\rm T}$ is the Thomson scattering cross-section, $G$ is the gravitational constant, $c$ is the speed of light, $\rho_{\rm gas}(r_{\rm A})$ is the gas density\footnote{Following \citet{valiante2011origin}, we assume the gas to follow an isothermal density distribution with a flat core. The core radius is assumed to be $0.012$ times the dark matter halo virial radius and the profile is normalized such that at each
time the total gas mass predicted by the model is enclosed within the virial radius. We refer the reader to \citet{valiante2011origin} for additional details.} at the radius of gravitational influence of the BH (also called Bondi radius, $r_{\rm A} = 2 G M_{\rm BH}/c_{\rm s}^2$), and $c_{\rm s}$ is the sound speed. The dimensionless parameter $\alpha_{\rm BH}$ is usually
introduced in numerical simulations to compensate for the lack of resolution that heavily underestimates the BHL accretion rate \citep[][]{springel2005modelling} and is a free parameter of the model (see Table \ref{tab:table1}).

Numerical simulations show that feedback due to SNe appears crucial in regulating BH growth in low-mass galaxies \citep{dubois2015} and therefore affects the BH mass distribution after the formation epoch of BH seeds \citep{habouzit2016}. Our model describes the total outflow rate from a galaxy as the sum of the active galactic nucleus (AGN) and SN contributions,
\begin{equation}
\frac{dM_{\rm ej}(t)}{dt}= \frac{dM_{\rm ej,AGN}}{dt}+\frac{dM_{\rm ej,SN}}{dt},
\label{eq:mej}
\end{equation}
\noindent
and the energy-driven AGN gas outflow (wind) is described as:
\begin{equation}
\frac{dM_{\rm ej,AGN}}{dt}= 2 \epsilon_{\rm w, AGN} \epsilon_{\rm r} \dot M_{\rm accr} (\frac{c}{v_{\rm e}})^2,
\label{eq:dmgrb}
\end{equation}
\noindent
where $\epsilon_{\rm w, AGN}$ is the wind efficiency. AGN winds are assumed to be in action when the accretion rate exceeds $10^{-2} \, \dot M _{\rm Edd}$ ($dM_{\rm ej,AGN}/dt = 0$ otherwise).
Even with our simplified treatment of AGN and SN feedback, BHs often starve due to gas evacuation, particularly in the small halos at high-$z$ that are characterized by small escape velocities. In this regime, the BH duty-cycle is set by halo mass growth via mergers, that provide additional gas supply to the nuclear regions, restarting a new cycle of star formation and BH growth.

\section{New features of the model} \label{section:newFeatures}

In this section we describe the new physically motivated prescriptions implemented in the model to investigate the occurrence of \textit{light}, \textit{heavy} and \textit{medium-weight} seeds in a cosmological context and their relative importance/role in the formation of $z>6$ SMBHs.

\subsection{Medium-weight seeds} \label{mwSeeds}
As we have seen in section \ref{section:BHseeds}, a necessary condition for the monolithic collapse is to avoid fragmentation. When dust-driven cooling
is effective, that is when $\mathcal{D} \ge \mathcal{D_{\rm cr}}$ \citep[][]{schneider2012}, fragmentation occurs at very high densities and in a very compact region, potentially leading to the formation of a dense stellar cluster \citep[][]{omukai2008can,clark2008first}. 

The question is then to understand what is the final fate of this system.
During their lifetime, collisional stellar systems evolve as a result of dynamical interactions. In an equal-mass system, the central cluster core initially contracts as the system attempts to reach a state of thermal equilibrium: energy conservation leads to a decrease in the core radius as evaporation of the less bound stars proceeds. As a result, the central density increases and the central relaxation time decreases. The core then decouples thermally from the outer region of the cluster. Core collapse then speeds up as it is driven by energy transfer from the central denser region \citep[][]{gurkan2004formation, reinoso2018,reinoso2019formation}. This phenomenon is greatly enhanced in multi-mass systems like realistic star clusters. In this case, the gravothermal collapse happens on a shorter time-scale as dynamical friction causes the more massive than $m_*$ stars to segregate in the center on a timescale $t_{\rm df} = (\langle m \rangle/m_*) \, t_{\rm rh}$, where $t_{\rm rh}$ is the half mass relaxation time-scale, and $\langle m \rangle$ is the mean stellar mass in the cluster \citep[][]{zwart2002runaway,devecchi2009formation}. If mass segregation sets in before the more massive stars evolve out of the main sequence ($\sim 3$ Myr), then a subsystem decoupled from the rest of the cluster can form, where star-star collisions can take place in a runaway fashion, ultimately leading to the growth of a very massive star (VMS). The upper limit on the duration of the process ($\sim 3$ Myr) is set by the requirement to avoid the explosion of the SNe, which would sweep away the system \citep[][]{sakurai2017formation}.

If a VMS forms above 260 $\rm M_{\odot}$, it would directly collapse into a BH of comparable mass.  In addition, the BH can still gain mass by accretion of stars and from the surrounding gas \citep[][]{devecchi2009formation}.  Since mass loss due to winds is significantly reduced in metal-poor stars, the growth of a VMS should be more efficient at low metallicity and the resulting
BH  mass can be as high as $10^3 \rm M_{\odot}$ \citep[][]{devecchi2009formation, volonteri2010formation}.

Hence, in \texttt{GQd} we assume that a \textit{medium-weight seed} forms in atomic cooling halos ($T_{\rm vir} \ge 10^4$\,K) where 
the gas is illuminated by a LW flux with $J_{\rm LW} \ge J_{\rm cr}$, the metal content of the ISM is subcritical, $Z < Z_{\rm cr}$, but the dust content is supercritical, $\mathcal{D} \geq \mathcal{D_{\rm cr}}$, so that dust cooling can lead to the formation of a dense star cluster. In addition, we require the halo to have a sufficiently large reservoir of gas, $M_{\rm gas} \geq 10^6 \,  \rm M_{\odot}$, to
form a dense and compact star cluster and we assume its final fate to be the
formation of a medium-weight seed with a mass of $10^3 \, \rm M_{\odot}$.

The environmental conditions required to form \textit{light, medium-weight}, and \textit{heavy} seeds in \texttt{GQd} are summarized in Table \ref{tab:table2}. In section \ref{section:results}, we discuss the effects of varying the adopted critical threshold for the LW flux to suppress H$_2$ formation. In addition, following \citealt{chon2020supermassive}, we also consider a variant of the reference model where we account for super-competitive accretion, that we illustrate below.

\subsection{The super-competitive accretion model}
\label{subsection:scamodel}
In a recent work \citet{chon2020supermassive} (see also \citealt{tagawa2020making}) used a suite of hydrodynamical simulations to investigate the outcome of star formation in slightly metal-enriched gas clouds illuminated by a very strong UV field. The initial conditions were set by  extracting -- from a cosmological simulation -- one halo that was shown to lead to the formation of a heavy seed when the gas was metal-free \citep[][]{chon2018}. As expected from previous semi-analytical models \citep[][]{omukai2008can}, 
dust-cooling becomes important when $Z \ge 5 \times 10^{-6} Z_{\odot}$ and multiple (about 600) low-mass fragments form in a relatively compact region with size smaller than 100 au. On larger scales, however, where dust cooling is not effective, the structure of the cloud is very similar to the primordial case: as a result of the high gas temperatures, matter continues to flow into the central regions, preferentially feeding the primary star (defined as
the first protostar that forms). In $10^4$ yr the primary star becomes supermassive, with a mass comparable to the
one found in the primordial case. Interestingly, a similar evolution is found for initial metallicities up to $Z = 10^{-4} Z_{\odot}$, although the initial number of low-mass proto-stars formed as a result of dust-cooling increases with $Z$ (about 4000 stars are formed when $Z = 10^{-4} Z_{\odot}$). When the metallicity becomes $Z = 10^{-3} Z_{\odot}$ (the maximum metallicity considered in their analysis), \citet{chon2020supermassive} found that metal fine-structure line cooling starts to be effective on larger scales (when the central gas density is $n \ge 10^4$ cm$^{-3}$), and the
associated decrease in the gas temperature leads to a suppression of the gas accretion rate by almost two orders of magnitude. As a result, the primary star mass grows only up to 350 $\rm M_{\odot}$. It is important to stress however, that
some of the secondary protostars merge with the primary stars. Although the fraction of mass acquired through
mergers is subdominant when $Z < 10^{-4} Z_\odot$, mergers contribute to more than 50\% of the final primary
mass when $Z = 10^{-3} Z_\odot$. Since the total stellar mass formed in this highest metallicity run is $\ge 10^3 \rm M_{\odot}$, this may imply that the resulting primary star may grow further if the evolution were followed for more than
the $10^4$ yr allowed by the simulation. 

Hence, in this super-competitive accretion (SCA) model, SMSs leading to heavy BH seeds may continue to
form at higher metallicities than assumed in our reference model. To investigate how this alternative scenario
would modify the relative importance of the different BH seed populations, we run additional simulations where
the conditions to form medium-weight and heavy seeds are set as follows:  atomic cooling halos illuminated by a super-critical LW flux host the formation of heavy seeds as long as $Z < 10^{-3.5} Z_\odot$, and of medium-weight seeds when $10^{-3.5} Z_\odot \le Z < 10^{-2.5} Z_\odot$ independent of their dust-to-gas mass ratio. Conversely, the transition from Pop III to Pop II stars is still assumed to occur when $Z \geq Z_{\rm cr} = 10^{-3.8} Z_{\odot}$ and $\mathcal{D} > 4.4 \times 10^{-9}$ so that this model variant does not directly affect the conditions to form light seeds (see Table \ref{tab:table2}).

\begin{table}
\caption{The environmental conditions adopted to plant the three families
of seeds in the \texttt{GQd} reference (R) model and in the super-competitive accretion (SCA) variant (see sections \ref{section:BHseeds} and \ref{section:newFeatures}). For each of these
models, two values of $J_{\rm cr}$ have been explored: 300 and 1000.}
\begin{tabular}{ccccc}
\hline
R model & $Z/Z_{\odot}$ & $\mathcal{D}$ & $J_{\rm LW}$ & $M_{\rm gas}$\\
\hline
light & $ <  10^{-3.8}$ & $< 4.4 \cdot 10^{-9}$ & $< J{\rm cr}$ & $> 0$ \\
medium & $<  10^{-3.8}$ &  $\geq 4.4 \cdot 10^{-9}$  & $\geq J{\rm cr}$ & $> 10^6 {\rm M_\odot}$\\
heavy & $<  10^{-3.8}$ &  $ < 4.4 \cdot 10^{-9}$ & $\geq J{\rm cr}$ & $> 10^6 {\rm M_\odot}$\\
\hline
\hline
SCA model & $Z/Z_{\odot}$ & $\mathcal{D}$ & $J_{\rm LW}$ &  $M_{\rm gas}$\\
\hline
light & $ <  10^{-3.8}$ & $< 4.4 \cdot 10^{-9}$ & $< J{\rm cr}$ & $ > 0$ \\
medium & $[10^{-3.5} - 10^{-2.5}]$ & all & $\geq J{\rm cr}$ & $> 10^6 {\rm M_\odot}$\\
heavy & $<  10^{-3.5}$ &  all  & $\geq J{\rm cr}$ & $> 10^6 {\rm M_\odot}$ \\
\hline
\end{tabular}
\label{tab:table2}
\end{table}

\subsection{Inhomogeneous feedback}
\label{section:chemevo}
In the \texttt{GQd} model, metals and dust returned to the ISM through SN explosions and AGB stellar winds are assumed to instantaneously mix in the ISM. In addition, SN- and AGN-driven outflows described in section
\ref{section:model} have a metal and dust composition that reflect the conditions at the outflow launching site. Once ejected in the IGM,
we assume metals and dust to uniformly mix with the surrounding gas. This is the pre-enriched medium that will be accreted onto newly virialized 
dark matter halos, and out of which new stellar and/or BH seeds will form.

In \citet{valiante2016first}, the formation rate of \textit{light} and \textit{heavy seeds}
was analyzed by applying the same conditions described in section \ref{section:BHseeds} and summarized 
in Table \ref{tab:table2} for the reference model. We found that, under the hypothesis of uniform metal/dust mixing and uniform LW background, the redshift window where halos satisfied the conditions for \textit{heavy seeds} formation was extremely 
narrow, and only a few \textit{heavy seeds} (between 3 and 30, depending on the simulation)
were able to form. This number, however, was enough to trigger the Eddington-limited growth
to a SMBH by $z \sim 6$ \citep[][]{valiante2016first}. 

However, metal (and dust) enrichment is known to be a highly inhomogeneous process, particularly at high redshift. 
Cosmological hydrodynamic simulations \citep[][]{tornatore2007population, maio2011, wise2012, johnson2013, muratov2013, pallottini2015, xu2016, jaacks2018, sarmento2019, graziani2020} supported by observations
\citep[][]{frederic2019, vanzella2020} show that metals ejected in the ISM by SNe and AGB stars propagate from 
higher density peaks to lower density regions, allowing the co-existence of pristine and metal enriched objects
in the pre-reionization epoch. 

Similarly, although the intensity of the LW background produced by stellar sources and accreting black holes 
increases very rapidly in the biased region where the bright quasar and its host galaxy are assembling 
(see Fig. 5 of \citealt{valiante2016first}), local fluctuations of the LW intensity from nearby sources
can significantly exceed the background level, suppressing H$_2$ cooling and aiding
to win the competition between LW radiation and metal winds in \textit{heavy seeds} hosts
\citep[][]{ahn2009,regan2014,dijkstra2014feedback,inayoshi2015suppression,habouzit2016,agarwal2017,maio2019, mayer2019}. 

Since we do not have any information on the spatial distribution of dark matter halos in the simulated volume, or on the spatial distribution of metal-enriched bubbles, we have implemented in \texttt{GQd} a statistical description of inhomogeneous chemical and radiative feedback that we present below.

\subsubsection{Inhomogeneous metal and dust enrichment} \label{sec:inomog}

Following \citet{dijkstra2014feedback} and \citet{salvadori2014}, at each redshift we compute the volume filling factor of the N enriched regions as:
\begin{equation}
    Q(z)=\sum_{\rm i=1} ^{\rm N} \frac{V_{\rm i}}{V_{\rm sim}}=\sum _{\rm i=1} ^{\rm N} \frac{4/3 \, \pi R^3_{\rm i}}{V_{\rm sim}},
\end{equation}
\noindent
where $\sum_{\rm i=1} ^{\rm N} V_{\rm i}$ is the volume filled by SN-driven metal-enriched bubbles with radius $R_{i}$ and $V_{\rm sim} = 50$ Mpc$^3 (1+z)^{-3}$ is the proper volume of the simulation, estimated at the turn-around radius. 
 At a given time (redshift), the radius of the $i$-th expanding bubble polluted by metals can be approximated by the self-similar
Sedov-Taylor blastwave solution \citep[][]{madau2001early,dijkstra2014feedback}:

\begin{equation}
    R_{\rm i}(t)= \xi_0 \, \biggl[\frac{E_{\rm SN}\, N_{\rm SN,i} }{\rho_{\rm gas}(z_{\rm i})} \biggr]^{1/5} t^{2/5},
    \label{eq:radius}
\end{equation}
\noindent
 where $\xi_0 = 1.17$, $E_{\rm SN} \, N_{\rm SN,i}$ is the total SN explosion energy in the i-th halo ($E_{\rm SN}$ is the averaged energy per SN, as in Eq. \ref{eq:mejsn}, and $N_{\rm SN}$ is the number of SNe)\footnote{ During the Sedov-Taylor expansion phase the energy is conserved and not radiated away. Hence, in Eq.\ref{eq:radius}, we consider all the energy (kinetic and thermal) injected by each SN.} $\rho_{\rm gas}$ is the gas density and $t = t(z)-t(z_{\rm i})$ is the interval between the current time $z$ and the redshift $z_{\rm i}$ at which the SN-driven outflow was launched. 
When the bubble is within the halo virial radius, $R_{\rm  i} < R_{\rm vir, i}$, $\rho_{\rm gas}$ is the mean gas density of the i-th halo at the moment of the SNe explosion. As the bubble expands beyond the halo virial radius into the IGM, $R_{\rm  i} > R_{\rm vir, i}$, we compute the gas density\footnote{ We note that this may lead to an unphysical jump in the gas density as the bubble reaches the virial radius. However, adopting a smoother transition, such as the one predicted by \citet{madau2001early}, delays the time evolution of the shell radius for a very small time interval and
provides a negligible correction to the time evolution of the filling factor.} as $\rho_{\rm gas}(z) = \Omega_{\rm b} \, \rho_{\rm cr}\,(1+z)^3$. 

At a given time (redshift) we define the mean metallicity, $Z_{\rm q}$, and dust-to-gas mass ratio, $\mathcal{D}_{\rm q}$ of the enriched regions as:
\begin{equation}
Z_{\rm q} (t) = \frac{M^{\rm tot}_{\rm met,ej}(t) }{M^{\rm tot}_{\rm ej} (t)+ [1-f_{\rm c}(t)]\, \sum_{\rm i=1}^{\rm N} \, \Omega_{\rm b} \, \rho_{\rm cr}(1+z)^3 V_i},
\label{eq:MET1}
\end{equation}
\noindent
and
\begin{equation}
\mathcal{D}_{\rm q}(t)= \frac{M^{\rm tot}_{\rm dust,ej}(t) }{M^{\rm tot}_{\rm ej}(t)+ [1-f_{\rm c}(t)]\sum_{\rm i=1}^{\rm N} \, \Omega_{\rm b} \, \rho_{\rm cr}(1+z)^3 V_i}.
\label{eq:dgratio}					 
\end{equation}
\noindent
where $M^{\rm tot}_{\rm ej}(t)$, $M^{\rm tot}_{\rm met,ej}(t)$ and $M^{\rm tot}_{\rm dust,ej}(t)$ are the total masses of gas, metals and dust that have been ejected out of the halos in the IGM, $[1-f_{\rm c}(t)]\, \sum_{\rm i=1}^{\rm N} \, \Omega_{\rm b} \, \rho_{\rm cr}(1+z)^3 V_i$ is the IGM gas mass enclosed in the volume covered by the bubbles expanding outside the halos (the summation runs on all halos for which the condition $R_{\rm  i} > R_{\rm vir, i}$ is satisfied), and $f_{\rm c}(t)$ is the time-dependent fraction of mass collapsed into halos.
The amount of gas ejected out of a single halo as a consequence of mechanical feedback, $M_{\rm ej}$, is computed as described in section~\ref{section:model}. The corresponding mass of ejected metals and dust are computed as $M_{\rm met,ej} = Z_{\rm ISM} M_{\rm ej}$ and $M_{\rm dust, ej} = {\cal D}_{\rm ISM} M_{\rm ej}$, where $Z_{\rm ISM}$ and ${\cal D}_{\rm ISM}$ are the metallicity and dust-to-gas ratio of the ISM \citep{valiante2011origin, valiante2014high}.

At each redshift, we assume that the \textit{i-th} halo has a probability $P_{\rm i}(z) \in [0,1]$, randomly extracted from a uniform distribution, to be located within a region of the IGM that has been (already) enriched by SN explosions (i.e. regions occupied by expanding bubbles). We then use the volume filling factor, $Q(z)$, to assign halos to the un-polluted or to the enriched volume: 
if $P_{\rm i}(z) > Q(z)$, the halo resides in the un-polluted volume $[1-Q(z)] \, V_{\rm sim}$, where it accretes gas of primordial composition; if $P_{\rm i}(z) \leq Q(z)$, the halo is instead associated to enriched regions, within the volume $Q(z) \, V_{\rm sim}$, and accretes polluted gas from the IGM\footnote{Note that, in the limit $Q(z) \rightarrow 1$, the IGM is completely filled by metal-enriched bubbles and the enriched volume converges to the total volume of the simulation: $\sum_{\rm i=1} ^{\rm N} V_{\rm i} \rightarrow V_{\rm sim}$.}.
At each given time, $Z_{\rm q}$ and $\mathcal{D}_{\rm q}$ define the abundance of metals and dust in the (infalling) material accreted by halos forming/evolving in enriched regions, as described above.
  \begin{figure}
     \hspace{-0.7 cm}
     \centering
     \includegraphics [scale=0.2]{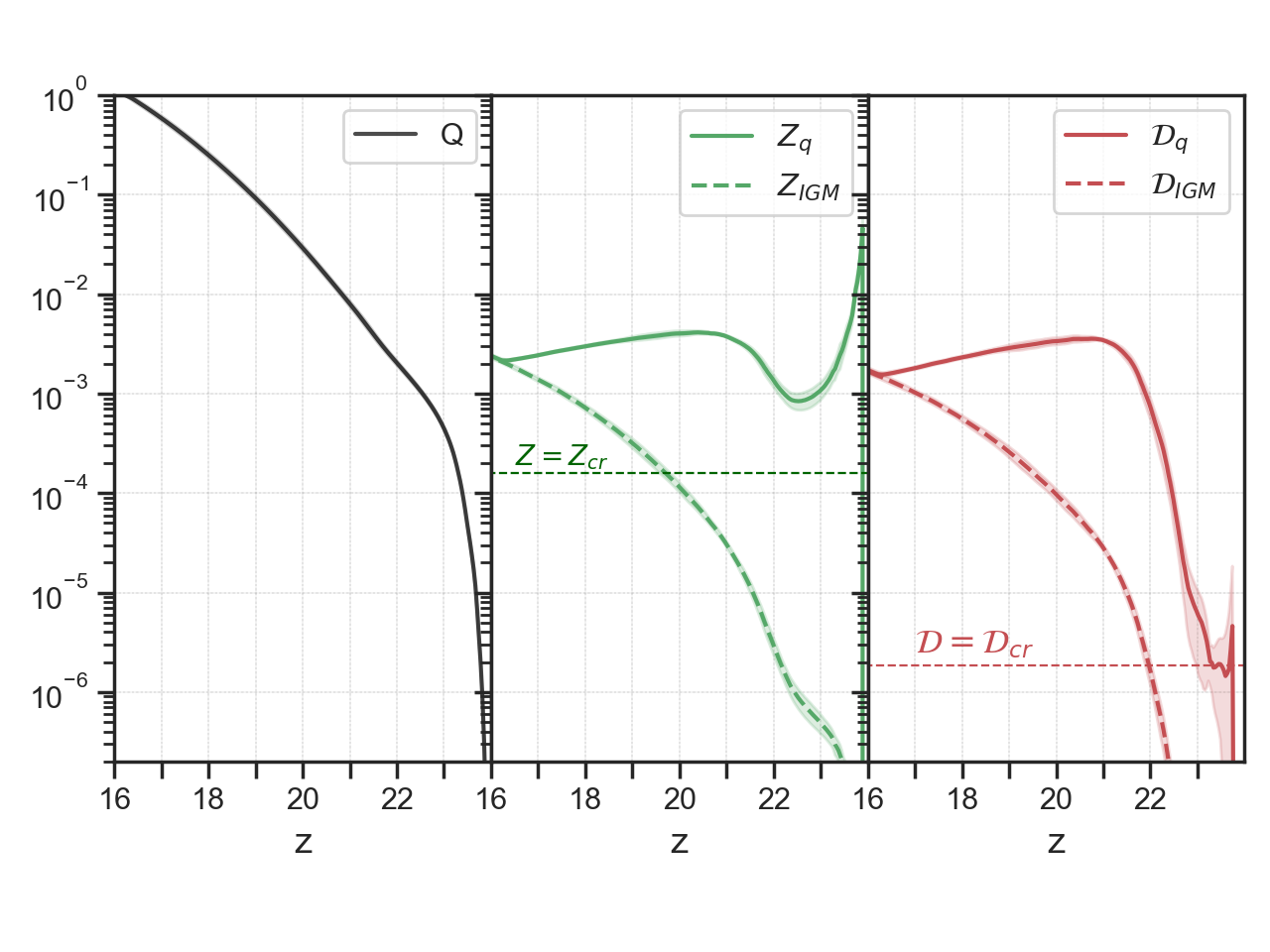}
     \caption{  
     Left panel: evolution of the filling factor of enriched regions (black line) as a function of redshift. Central panel: mean metallicity of the IGM (green dashed line) and of the enriched regions (green solid line) expressed in solar units, with $Z_\odot = 0.02$. Right panel: mean dust-to-gas ratio of the IGM (red dashed line) and of the enriched regions (red solid line) normalized to the Galactic value, ${\cal D}_{\rm MW} = 0.006$. The lines are mean values over ten realizations of the reference model, with 1- $\sigma$ errorbars (light shaded areas). Horizontal dashed lines in the central and right panels are the critical values for metallicity ($Z_{\rm cr}$) and dust to gas ratio ($\mathcal{D}_{\rm cr}$) adopted in the reference model. 
        }
     \label{fig:zqd}
 \end{figure}
Figure \ref{fig:zqd} shows the redshift evolution of the filling factor ($Q$, left-hand panel), 
metallicity ($Z_{\rm q}$, central panel), and dust-to-gas mass ratio (${\cal D}_{\rm q}$, right-hand panel) 
of enriched regions predicted by the reference model (see section \ref{section:results}). While $Q$ monotonically increases with cosmic time, reaching the value $Q = 1$ at $z \sim 16$, the redshift evolution of $Z_{\rm q}$ and ${\cal D}_{\rm q}$ is not monotonic and depends on the 
metal and dust yields of the dominant SN population, on their energies, and on the mass of gas swept by the
expanding bubbles. At early times, when the SFR is dominated by Pop III stars, the bubbles
are still relatively small ($Q << 1$) and there is a very rapid increase in both 
$Z_{\rm q}$ and ${\cal D}_{\rm q}$, implying that the few enriched regions are very metal- and dust-rich. 
As the bubbles expand, progressively enclosing a larger IGM gas mass, $Z_{\rm q}$ and ${\cal D}_{\rm q}$ 
decrease. When the star formation rate starts to be dominated by Pop II stars, at $z \lesssim 23$,
$Z_{\rm q}$ and ${\cal D}_{\rm q}$ increase again but then smoothly decline until, at $z \sim 16$, when $Q \sim 1$, 
the volume is homogeneously enriched and their values converge to the mean IGM metallicity, $Z_{\rm IGM}$, and dust-to-gas mass ratio, ${\cal D}_{\rm IGM}$ (shown in Fig. \ref{fig:zqd} as dashed lines). 
The two horizontal lines in Fig. \ref{fig:zqd} represent the adopted critical metallicity and dust-to-gas ratio
in the reference model (see Table \ref{tab:table2}). The figure shows that the condition $Z_{\rm IGM} \sim Z_{\rm cr}$ 
(${\cal D}_{\rm IGM} \sim {\cal D}_{\rm cr}$) is met already at $z \sim 20$ ($z \sim 22$). These are the reshifts 
below which BH seeds can no longer form in models where the IGM is assumed to be uniformly enriched. However, at these
cosmic epochs, we predict the filling factor to be still very small ($Q \sim 10^{-2}$) and metal enrichment to be
highly inhomogeneous. As a consequence, BH seeds can continue to form down to $z \sim 16$, when the volume has been completely filled by metal-enriched bubbles ($Q = 1$).  Hence, inhomogeneous enrichment extends the redshift range where BH seeds may potentially form from $z \sim 20$ to $z \sim 16$.

As a final remark, we notice that the above model of inhomogeneous enrichment assumes that the individual enriched bubbles do not overlap. This neglects the effects of clustering, which are likely to be significant for high-redshift star-forming galaxies and may lead to a $\sim 70\%$ reduction of $Q$ at $z < 17$, assuming that the enrichment is mostly driven by $10^6 - 10^7 \, M_\odot$ halos\footnote{The average comoving distance between $10^6 - 10^7 \, M_\odot$ halos at $z = 10 - 20$ ranges between $<d> \sim 0.13$ to $\sim 1.1$\, Mpc. The excess probability of finding two halos with mass $M = 10^6 M_\odot$  ($10^7 M_\odot$) separated by a comoving distance $<d>$ at $10 < z < 20$ can be estimated to range between 1 and 3 (see e.g. Fig. 2 in \citealt{iliev2003non}). Assuming that within $<d>$ we find 3 times more systems than if these were uniformly distributed, and that their individual metal-enriched bubbles overlap, our estimated filling factor should be reduced by $ \sim 70\%$ to account for this clustering effect.}. The effects of this reduction would be to extend the epoch of BH seed formation to lower redshift and to favour the formation of heavy BH seeds, that preferentially form in metal-free atomic-cooling halos illuminated by a super-critical LW flux, at the expenses of medium-weight BH seeds (see section \ref{section:BHseeds}). However, this does not significantly affect the resulting evolution of the SMBH mass (see section \ref{section:BHgrowth}).

\subsubsection{Inhomogeneous $J_{LW}$ flux}

Inhomogeneities in the LW flux and their relation with potential sites for heavy BH seeds formation have been investigated for many years \citep[][]{dijkstra2008fluctuations, ahn2009, agarwal2012ubiquitous, dijkstra2014feedback,  inayoshi2015suppression}.
In fact, in the redshift range where BH seeds are expected to form, the mean LW background is significantly
smaller than the required critical value above which H$_2$ cooling in atomic cooling halos is suppressed,
unless the region under investigation is particularly over-dense \citep[][]{valiante2016understanding,valiante2016first}.
Hence, heavy seeds form when their potential host halo is illuminated by a super-critical $J_{\rm LW}$ from a nearby
galaxy \citep[][]{visbal2014, habouzit2016number, regan2019emergence, maio2019}, whose distance must not be too close in order to avoid gas stripping or tidal distortions, which would prevent the formation of the SMS \citep[][]{chon2016}.

This motivates the importance of considering the non-linear spatial distribution of the UV emitting sources around a certain atomic cooling halo, $M_1$, candidate to host a heavy or medium-weight seed. In other words, the proximity of $M_1$ to a star-forming halo can easily help the halo to meet the condition that $J_{\rm LW} \geq J_{\rm cr}$.

To establish the distance between $M_1$ and a UV source, we start by considering the probability to find another halo at distance $r$. Following \citet{inayoshi2015suppression}, the differential probability distribution of finding a halo with mass 
$M_2 \pm dM/2$ at a distance $r \pm dr$ from $M_1$ is given by
\begin{equation}
\frac{d^2P(M_1,M_2,z,r)}{dM dr}= 4 \pi r^2(1+z)^3 \times
 [1+\xi(M_1,M_2,z,r)]\frac{dn}{dM},
\end{equation}
\noindent
where $dn/dM$ is the comoving number density of $M_2$ halos at redshift $z$ and $\xi$ is the non-linear bias \citep[][]{iliev2003non}.

At each redshift of the merger tree, we identify -- among all simulated halos -- those candidate to host a BH seed (see section \ref{section:BHseeds} and Table \ref{tab:table2}), $M_1$. For each of these, we consider $N_r$ spherical shells centered on $M_1$, with radii $r_i$
varying from $R_{\rm min}=10^{-3}$ Mpc to $R_{\rm max}= (3 V_{\rm sim}/4\pi)^{1/3}$ Mpc. We then randomly sample the cumulative probability to find,
among all possible star-forming halos $N_s$, a source $j$ at distance $r_i$ and we compute the LW intensity that it emits due to star formation, $I^j_{\rm stars}$ and BH accretion $I^j_{\rm AGN}$. Finally, we sum all the contributions to compute the total LW flux irradiating each $M_1$:
\begin{equation}
    J_{\rm LW}=\sum _{j=1}^{N_{s}} \sum _{i=1}^{N_{r}} \frac{(I_{\rm stars,i}^{j}+I_{\rm AGN,i}^{j}) }{4 \pi  r_{i}^2}.
\end{equation}
\noindent
For the present study, we follow \citet{iliev2003non} and adopt $N_r=300$. 
$I^j_{\rm stars}$ is computed using the mass (or age) and metallicity-dependent emissivities by \citet{schaerer2002} (for Pop~III stars) and \citet{bruzual2003} (for Pop~II/I stars) while $I^j_{\rm AGN}$ is obtained from a standard multi-temperature disc model with the addition of a non-thermal power-law component at high energies \citep[$\propto \nu^{-2}$][]{shakura1973, sazonov2004}.

\begin{figure*}
     \centering
     \includegraphics [width=11cm]{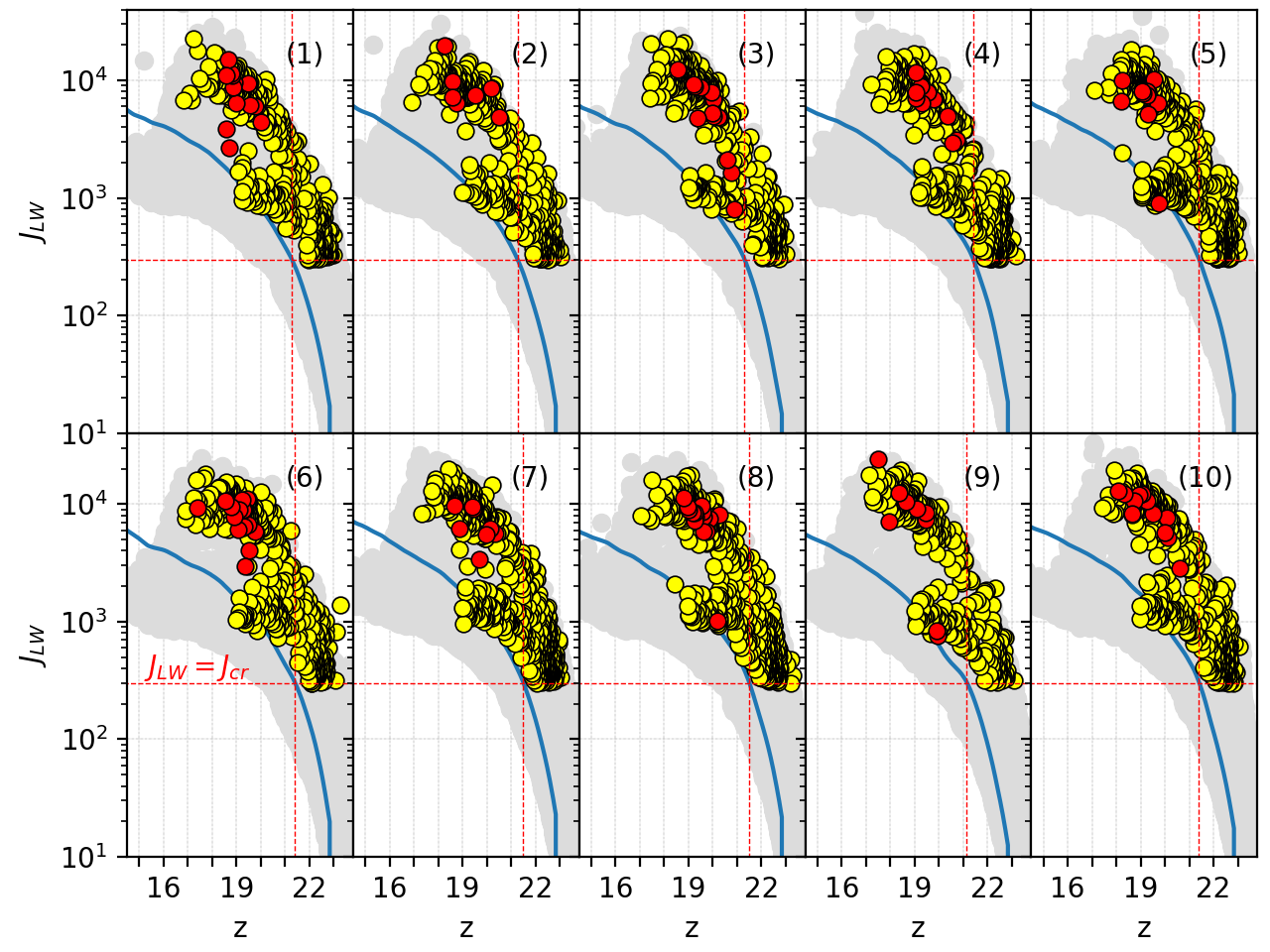}
     \caption{The intensity of the LW radiation (in units of $10^{-21}$ erg/s/Hz/cm$^2$/sr) as a function of $z$ for ten different simulations of model R300. Each grey point represents a single halo and we have highlighted in red and yellow the formation sites of heavy and medium-weight seeds respectively. The solid blue line is the mean LW background and the red dotted lines identify the redshift at which the mean LW background exceeds $J_{\rm cr} = 300$.}
     \label{fig:jlw}
 \end{figure*}

\section{Results}
\label{section:results}

In this section, we start by illustrating the results of the reference model, R300, where we adopt a critical value for the LW flux of $J_{\rm cr} = 300$. We first present the properties of BH seeds formation environments, such as the intensity of the illuminating LW radiation, $J_{\rm LW}$, the metallicity, and dust-to-gas mass ratio.
Then, we quantify the relative contribution of different seed populations to the mass growth history
of the final SMBH. We finally explore the implications of changing the value of $J_{\rm cr}$ and how the growth
history of the SMBH is modified when the SCA model is adopted.

\subsection{Mean and local LW flux}

In Fig. \ref{fig:jlw}, we show the predicted evolution of the mean LW background as a function of redshift (blue lines) in 10 independent simulations of R300. 
 It is important to stress that this quantifies the LW background in our simulated volume that represents a very biased region that will form a massive halo at $z = 6.4$. This is why the solid lines reach values that are larger than
the cosmic averages typically found in other studies (see, for instance, Fig. 6 in \citealt{ahn2009} or Fig. 3 in \citealt{dijkstra2014feedback} where their mean LW background at $z > 10$ is found to be in the range $\sim 0.1 - 10$). Since the mean LW background is roughly proportional to the stellar mass density (see Eqs. 12 and 13 in \citealt{agarwal2012ubiquitous}), a larger background is to be expected in the highly biased region that will assemble a $10^{13} M_\odot$ halo at $z = 6.4$. Indeed, \citet{petri2012} find comparably large values when they simulate the assembly of a $10^{12} M_\odot$ halo at $z \sim 6$ (see their Fig. 5). 

Conversely, our predicted local fluctuations in the LW field are very consistent with previous findings (see, for example, Fig. 3 of \citealt{agarwal2012ubiquitous} or the colour map in Fig. 1 of \citealt{dijkstra2014feedback}). These are shown with grey points in Figure \ref{fig:jlw},  that represent the (local) intensity of LW radiation illuminating individual halos. Among these, we have marked in red (yellow) the formation sites of heavy (medium-weight) seeds. Red dotted lines identify the redshift 
at which the mean LW background exceeds the critical value $J_{\rm cr} = 300$. In all simulations, this condition is met at $z \sim 21$.  While medium-weight seeds can form above this redshift, we find that there are large fluctuations around the mean LW background 
and in most cases heavy and medium-weight seeds form in environments where $J_{\rm LW} > 10^3$, largely exceeding our adopted value for $J_{\rm cr}$.

\begin{figure}
     \hspace{-0.7 cm}
     \centering
     \includegraphics [width=9 cm]{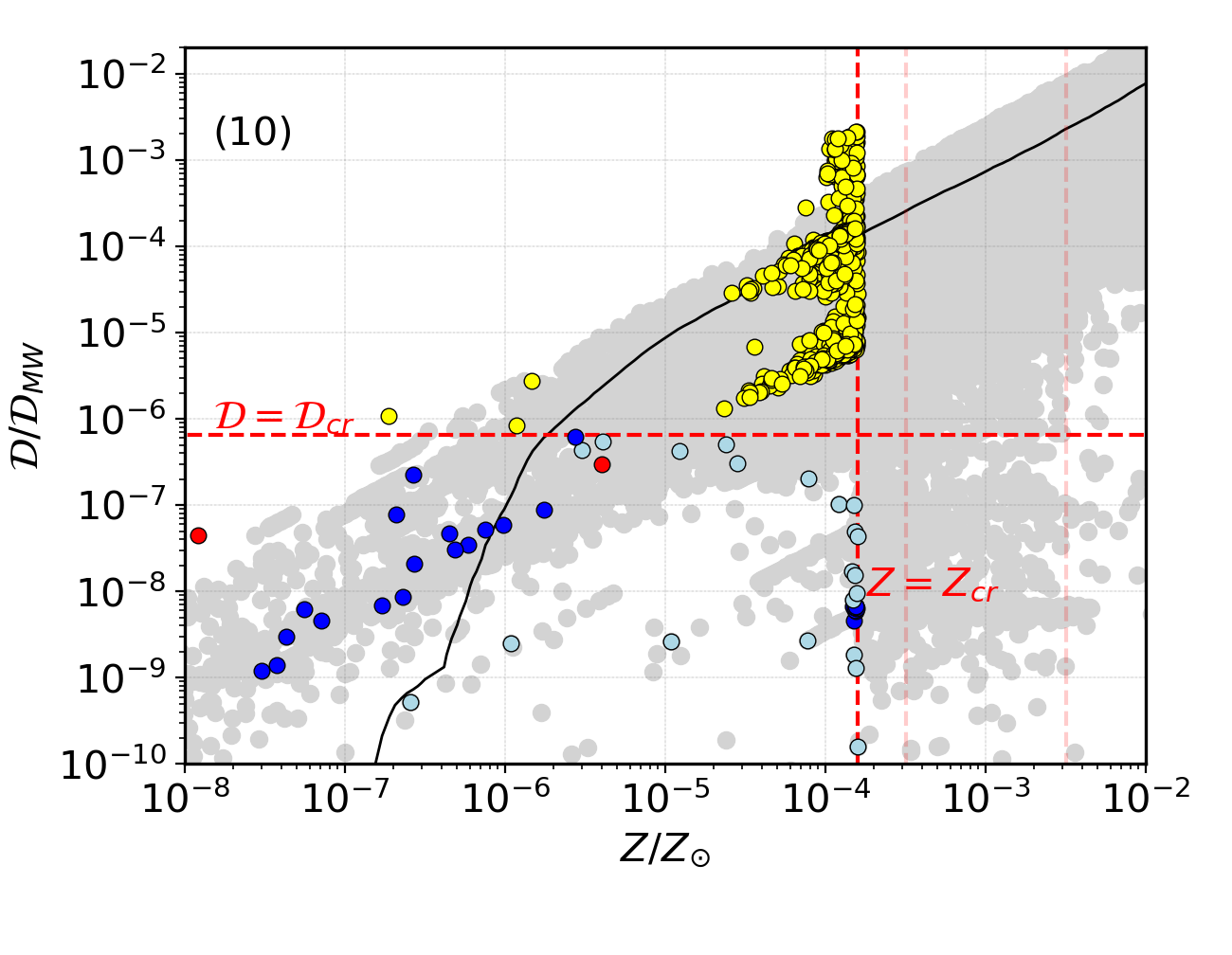}
     \caption{The ISM dust-to-gas mass ratio and metallicity of all the star-forming galaxies in simulation (10) at $z \ge 16$. Each grey point represents a single halo and - similarly to Fig.\ref{fig:jlw} - we have marked in red and yellow the formation sites of heavy
     and medium-weight BH seeds. We note that the majority of heavy seeds form in pristine galaxies that are not accounted for in this plane. Cyan and blue points indicate, respectively, non-pristine minihalos and atomic-cooling halos where light seeds form. The black solid line represents the mean values of the IGM. Red dashed lines are the critical dust-to-gas ratio and metallicity thresholds adopted in the reference model. For comparison, the two thin vertical lines show the critical metallicity thresholds adopted in the SCA model (see Fig. \ref{fig:zvsdustSCA} and Table \ref{tab:table2})}.  
     \label{fig:zvsdust}
 \end{figure}

\subsection{Mean and local metal enrichment}

In Fig.~\ref{fig:zvsdust}, we report the ISM metallicity and dust-to-gas mass ratios of individual
galaxies extracted from a single simulation (simulation 10 shown in the bottom-right panel of Fig.~\ref{fig:jlw}) 
from the onset of chemical enrichment down to the redshift at which the filling factor of enriched 
regions, $Q$, is equal to 1 ($z \ge 16$). We follow the same colour code adopted in Fig.~\ref{fig:jlw}
and we mark in red and yellow systems hosting heavy and medium-weight BH seeds, respectively. 
In addition, we also show the formation sites of light BH seeds, marking in cyan and blue systems hosted, 
respectively, in minihalos and atomic cooling halos.
The solid black line indicates the mean values of the IGM, that shows a monotonic increase of ${\cal D}_{\rm IGM}$ with $Z_{\rm IGM}$\footnote{We point out that ${\cal D}$ and $Z$ are not always proportional because of dust reprocessing in the ISM due to SN dust destruction and grain growth (see section \ref{section:chemevo}).}.  As expected from Fig.~\ref{fig:zqd}, when $Q << 1$, galaxies located within the already enriched regions have a metallicity and dust-to-gas mass ratio that largely exceed the mean IGM value (and above the critical metallicity and dust-to-gas mass ratio thresholds) populating the upper-right end of Fig.~\ref{fig:zvsdust}, together with their lower-redshift and metal/dust-rich descendants. On the other hand, unpolluted regions at $z>16$ host metal/dust-free ($Z=\mathcal{D}=0$) galaxies that do not appear in the figure. This is why the density of grey points is significantly smaller in the bottom-left part of the plane.

The two red dashed lines mark the critical values for $Z_{\rm cr}$ and $\mathcal{D}_{\rm cr}$ 
adopted in the reference model. 
These divide the plane in four separate regions: 
the bottom-left part contains the most pristine environments, where only light and
heavy BH seeds form. We  note  that  the  number  of  red  points  is significantly smaller than  Fig. \ref{fig:jlw} (see the bottom-right panel).
In fact, most of the heavy BH seeds form in pristine halos, that are not visible in Fig. \ref{fig:zvsdust}.

A similar consideration applies to light BH seeds, a large fraction of which form when $Z = {\cal D}=0$.
In the top-left region only medium-weight seeds can form, in the sub-sample of halos that are illuminated by a super-critical $J_{\rm LW}$ and that contain a sufficiently large gas mass ($\ge 10^6 \rm M_{\odot}$). The latter condition preferentially selects systems along diagonal lines on this plane. The highest density of galaxies appears
to the right of the vertical line representing $Z = Z_{\rm cr}$, where Pop II/I stars are assumed to
form by metal and dust-driven fragmentation. Indeed, their parent star-forming clouds are either
initially enriched above $\mathcal{D}_{\rm cr}$ (top-right region) or they reach this condition during 
cloud collapse through grain growth \citep[][]{chiaki2014, chiaki2015}.

\begin{figure}
     \hspace{-1.18 cm}
     \centering
     \includegraphics [scale=0.185]{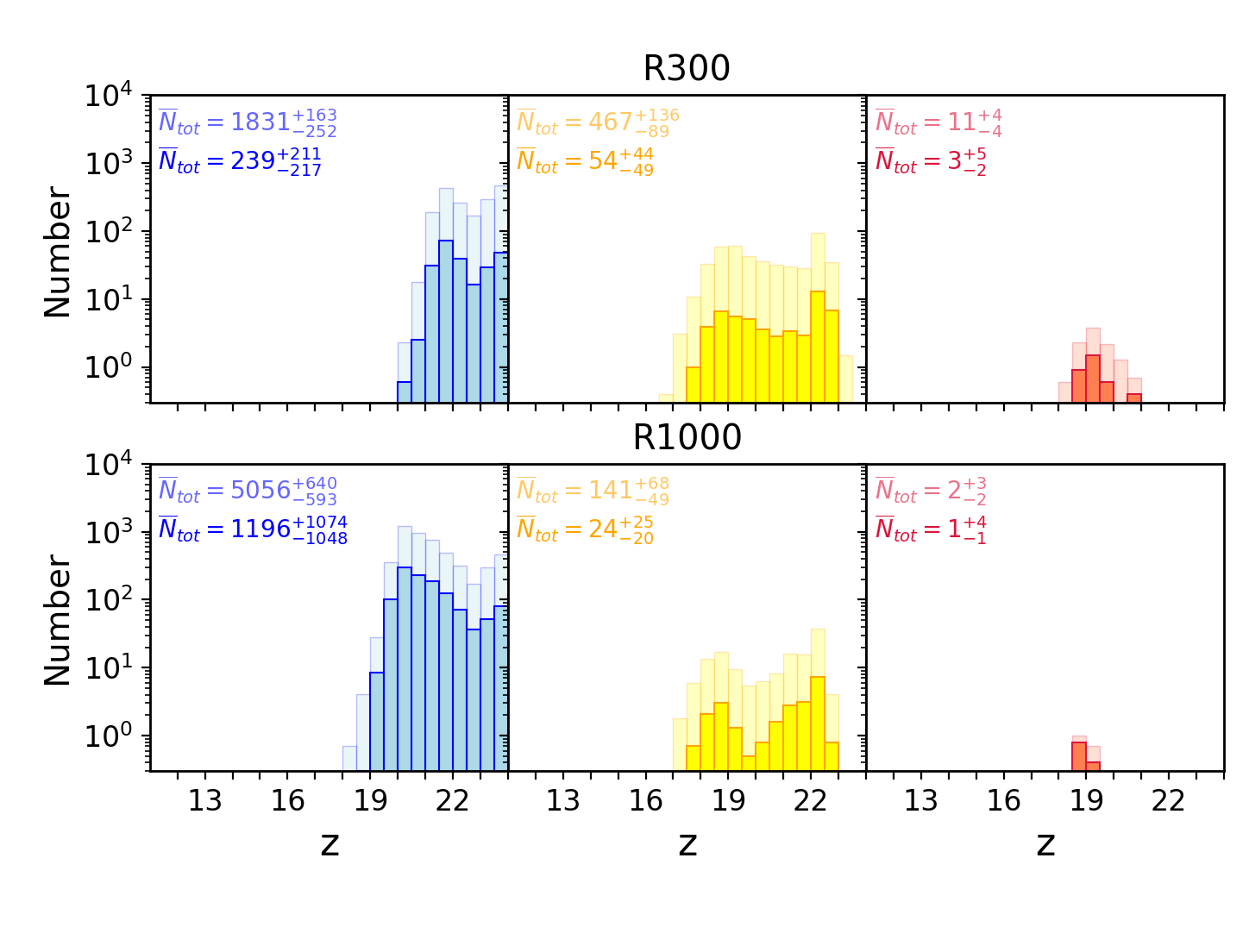}
     \caption{
     Distribution of formation redshifts of light, medium-weight, and heavy BH seeds (from left to right). The  histograms indicate
     the mean values averaged over 10 different simulations. Histograms in lighter colours show the global BH seed population, whereas histograms in heavier colours illustrate the true progenitors of the final SMBH at $z=6.4$ (see text). In each panel, we also indicate the total number of seeds formed, on average, in the corresponding population. The errors indicate the difference with the maximum and minimum values for each family over 10 simulations. 
     We compare the results for the same reference set of parameters, assuming $J_{\rm cr} = 300$ (model R300, top panels) and $J_{\rm cr} = 1000$  (model R1000, bottom panels).}
     \label{fig:Rseeds}
 \end{figure}

\subsection{Seeds birth environments}

In the top panel of Figure \ref{fig:Rseeds}, we show the distribution of formation redshifts of light, medium-weight, and heavy BH seeds for model R300. The bottom panel will be discussed in section \ref{subsec:Jcr}. 
The light-coloured histograms represent the averaged values found in 10 independent simulations and in each panel we also report the corresponding total number of BH seeds formed. Histograms in heavier colours illustrate the sub-sample of each family of seeds that are the progenitors of the final SMBH at $z=6.4$ (those systems that directly contribute to its mass assembly\footnote{As explained in section \ref{section:BHgrowth}, the progenitors of the final SMBH are seed BHs that - while growing - remain in the nuclei of their galaxies and never become wandering BHs as a consequence of minor mergers experienced by their host halo.}). The errors associated to the number of BH seeds (and BH seeds progenitors) correspond to the difference with the maximum and minimum values found in the 10 simulations. 

We find that the epoch of formation of BH seeds starts at $z \sim 24$, when H$_2$ cooling in the first mini-halos becomes efficient enough to trigger star formation\footnote{According to \citet{valiante2016first}, the fraction of metal-free gas that is able to cool in one dynamical time, $f_{\rm cool}$ (see Eq. 1), rapidly drops to zero when $T_{\rm vir} < 2000$\, K at $z \geq 25$, even when $J_{\rm LW} = 0$ (see the top left panel in Fig. A1 of \citealt{valiante2016first}). This limit corresponds to a minimum halo mass that can host star formation of $\sim 10^6 \, \rm M_\odot$ at $z \sim 24$, consistent with our resolution mass in the merger tree simulations.}. As expected, light BH seeds form first and largely outnumber medium-weight
and heavy BH seeds. On average, we find $\sim 1831$ light BH seeds forming at $z \ge 20$, while $\sim 467$ medium-weight and $\sim 11$ heavy BH seeds form at $16 \le z \le 23.5$, when atomic-cooling halos start to assemble. Termination of light BH seed formation is set by radiative feedback, as at $z \le 20$ metal-poor or pristine star-forming regions are illuminated by a super-critical LW flux (see Fig. \ref{fig:jlw}) and meet the condition to form medium-weight and heavy BH seeds. We note that medium-weight seeds start to form before their heavy counterparts. Indeed,
while the latter preferentially form in pristine halos, the former can seed halos that have already
experienced previous episodes of star formation but whose $Z$ and $\mathcal{D}$ still meet the conditions
required to form medium-weight seeds. 

The above results can be compared to \citet[][see in particular their Fig. 4]{valiante2016first}. 
While the results of the above study are qualitatively similar, our inhomogeneous treatment of chemical and radiative feedback leads to an increase in the absolute number of light seeds ($\sim 830$ light and $\sim 10$ heavy BH seeds were predicted to form, on average, in \citealt{valiante2016first}) and to a more uniform distribution in the formation epoch of heavy BH seeds (the majority of which where expected to form at $z \sim 16$ in \citealt{valiante2016first}), thanks to the higher number of pristine regions and to the fluctuations in the LW flux.

The comparison between lighter and darker histograms in each panel shows that the formation redshift distributions of BH seeds progenitors follow the same trend of their parent populations but with smaller numbers: on average, only $\sim 12-13$ per cent of medium-weight and light and $\sim 27$ per cent of heavy BH seeds are progenitors of the final SMBH.

 As a final remark, we stress that these absolute numbers quantify the seeds populations within a comoving volume of 50 Mpc$^3$ selected to host a $10^{13} \rm M_{\odot}$ dark matter halo at $z = 6.4$. Hence, it is not straightforward to interpret these numbers as representative of an average cosmic region of the Universe. Rather, they should be considered as the expected BH population along the formation route of each $z > 6$ SMBH.

 \begin{figure}
     \hspace{-1.15 cm}
     \includegraphics [scale=0.25]
     {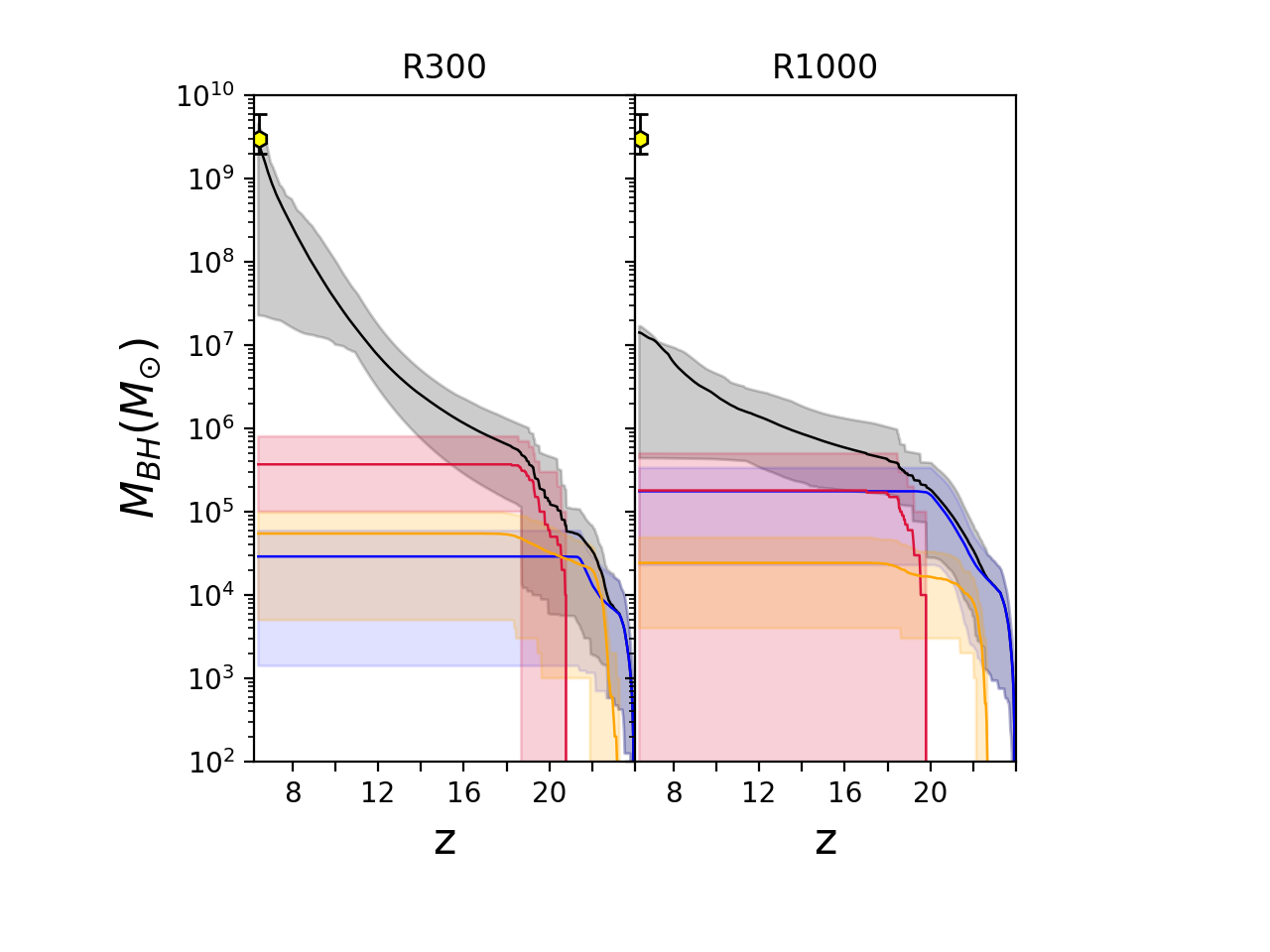}
     \caption{Evolution of the total mass of nuclear black holes as a function of redshift (black line) with the separate contributions of light (blue line), medium-weight (yellow line), and heavy BH seed (red line) progenitors. 
     Each line represents the average value among 10 simulations, with the shaded region ranging between the minimum and maximum value found at each redshift. 
     We compare the results for the same reference set of parameters but assuming $J_{\rm cr} = 300$ (model R300, left panel) and $J_{\rm cr} = 1000$  (model R1000, right panel). The data point at $z = 6.4$ represents the estimated SMBH mass of the QSO $J1148$ that we have assumed as a prototypical system and against which we have calibrated the model free parameters.}
     \label{fig:bhevoR}
 \end{figure}

\subsection{Nuclear BH mass evolution}
\label{subsec:Nuclearbh}

In the left panel of Fig. \ref{fig:bhevoR}, we show the resulting evolution of the nuclear BH mass as a function of redshift for model R300. 
This is obtained by summing over all BH seed progenitors masses\footnote{It is important to stress that we restrict the analysis to real SMBH progenitors, meaning that we reconstruct the mass and redshift distributions of all
the nuclear BHs that contribute to the final SMBH mass.} present at each redshift in the nuclei of the simulated galaxies, starting from the multiplicity of small systems hosted in minihalos at $z = 24$, 
down to the single $10^{13} \rm M_{\odot}$ halo at $z = 6.4$. The figure shows the average evolution found among 10 different simulations considering accretion (black line) and illustrates the separate contributions to the 
total mass growth of light (blue line), medium-weight (yellow line), and heavy (red line) BH seed progenitors,  without
taking into account their subsequent growth by gas accretion. For each of these lines, the shaded area with the same colour ranges between the minimum and maximum values found at each redshift. 

As expected from Fig. \ref{fig:Rseeds}, the sequence of coloured lines reflects the formation redshifts of the three BH seeds
populations. Despite their smallest number, the greatest contribution to the mass growth is provided
by heavy seeds, (with a cumulative BH mass of $\sim 3 \times 10^5 \, \rm M_{\odot}$),  that is much larger than those of
light ($\sim 3 \times 10^4 \rm \, M_{\odot}$) and medium-weight seeds ($\sim 5 \times 10^4 \, \rm M_{\odot}$). At $z < 16$, BH seeds no longer form and the growth of nuclear BHs is entirely driven by gas accretion, which provides the dominant contribution
to the final SMBH mass of $\sim 3 \times 10^9 \rm M_{\odot}$.

The above findings appear to be in very good qualitative agreement with the results obtained by \citet[][see in particular their Fig. 3]{valiante2016first}, where they also find that the Eddington-limited growth of $z \sim 6$ SMBHs relies on the formation of a small number of heavy seeds (with a total BH mass of $\sim 10^6 \, \rm M_{\odot}$) at early times and is dominated by gas accretion at late times. While the present study confirms this qualitative picture, it also shows that medium-weight BH seeds appear to be more frequent than heavy seeds among the ancestors of $z \sim 6$ SMBHs but - due to their 
smaller BH masses - their contribution to the total mass growth is largely subdominant.

\subsection{Dependence on the critical LW flux}
\label{subsec:Jcr}

The results presented above have been obtained selecting the environmental conditions to plant the three families of BH seeds according to a set of parameters that characterizes the reference model (see Table \ref{tab:table2}).  Here we wish to explore the sensitivity of our results to the adopted critical value of $J_{\rm cr}$. Indeed, the critical value of the LW flux for heavy (and medium weight) BH seed formation is admittedly very uncertain. The recent review by \citet{inayoshi2020} provides a thorough discussion of the various results that have been found by different groups using one-zone models or full 3D simulations. In general, the value of $J_{\rm cr}$ depends on the details of the calculation of the optically-thick H$_2$ photo-dissociation rate, which is challenging even in one-zone models. Using emissivity properties that are realistic for low-metallicity galaxies, \citet{sugimura2014critical} have provided the most complete calculations, finding $J_{\rm cr}$ to vary in the range $\sim  1000 - 1400$, with the lower end that is generally favoured when self-shielding and non-LTE effects are considered \citep{wolcottgreen2017, wolcottgreen2020suppression}. Full 3D simulations, that account for dynamical effects but generally implement simplified treatments of the shielding factor, tend to find values of $J_{\rm cr}$ that are a factor of few higher than in one-zone models (see \citealt{inayoshi2020} and references therein).

\begin{figure}
\hspace{-0.8 cm}
     \centering
     \includegraphics [scale=0.2]{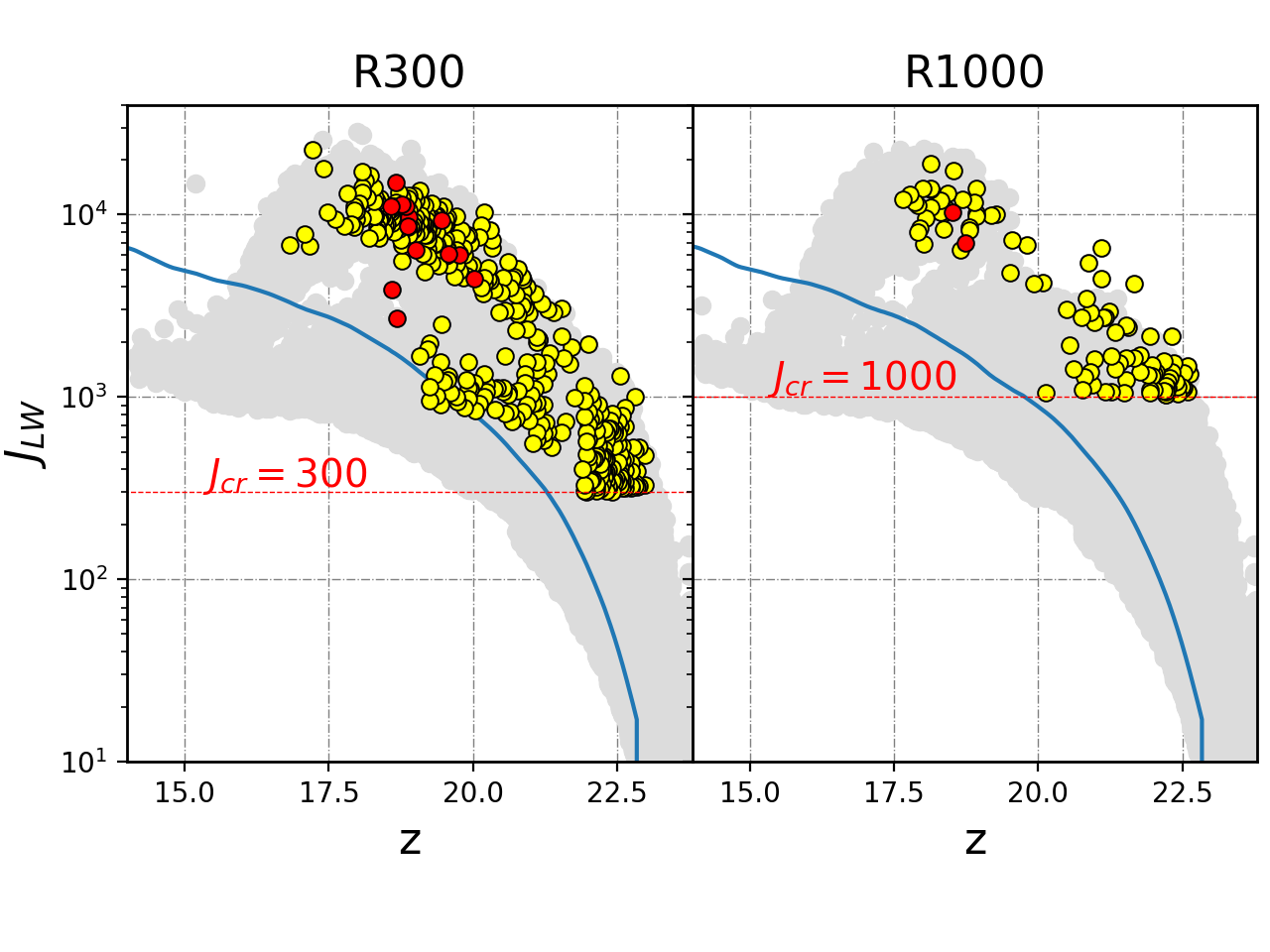}
     \caption{Same as Fig. \ref{fig:jlw} but for a single simulation and comparing the results of the reference model with $J_{\rm cr} = 300$ (model R300, left panel, i.e. run 1 of Fig. \ref{fig:jlw}) and $J_{\rm cr} = 1000$ (model R1000, right panel).}
     \label{fig:stdJ21}
 \end{figure}

To explore the impact of this parameter on our results, we have run a set of new simulations of the reference model adopting 
the increased value of $J_{\rm cr} = 1000$ (model R1000), as suggested by \citet{sugimura2014critical}. A comparison between model R1000 and the reference simulation (model R300) is provided in Figs. \ref{fig:Rseeds}, \ref{fig:bhevoR}, and  \ref{fig:stdJ21}.

A major difference between the two models is the decrease in the number of medium-weight and heavy BH seeds in model R1000 (by a factor of $\sim 3.3-5.5$), partly compensated by a comparable increase of light BH seeds (see Fig. \ref{fig:Rseeds}). Indeed, a larger $J_{\rm cr}$ implies a less effective radiative feedback and that star
formation in metal-free or metal-poor systems, that rely on H$_2$ cooling, is no longer suppressed (i.e. all systems
exposed to $300 < J_{\rm LW} < 1000$ can form stars in R1000). Yet, due to the fluctuations of the LW radiation, 
halos that host the formation of a medium-weight or heavy BH seeds can be illuminated by a LW flux that 
largely exceeds $J_{\rm cr}$. This is clearly shown in Fig. \ref{fig:stdJ21}, that is the analogous of Fig. \ref{fig:jlw},
but comparing the results of a single simulation in models R300 (left panel) and R1000 (right panel). In model R300 a large fraction of medium-weight seeds and all heavy BH seeds are predicted to form in systems illuminated by a LW intensity
$> 2000$. Hence, their formation should only be mildly affected by an increase in $J_{\rm cr}$ from 300 to 1000. However,
the strong interplay between radiative and chemical feedback at these redshifts is such that the less effective
radiative feedback leads to an increase in the (Pop III) SFR and to a more efficient metal and dust enrichment at these epochs. Due to star formation in their progenitors, most of the halos that host the formation of a medium-weight or heavy BH seed in model R300 are already too enriched in model R1000 and no longer meet the conditions to form a BH seed. 

Similar considerations apply if we restrict the analysis to the population of BH seed progenitors, represented by the darker coloured histograms in the bottom panels of Fig. \ref{fig:Rseeds}. The increased value of $J_{\rm cr}$ in model R1000 leads to an increase by a factor $\sim 5$ in the number of light BH seed progenitors and to
a factor $\sim 2.2-3$ reduction in the number of medium-weight and heavy BH seed progenitors
with respect to model R300. As a result, Figure \ref{fig:bhevoR} shows that the Eddington-limited growth of the BH seed progenitors formed in model R1000 leads to the formation of a $z \sim 6$ SMBH with an average mass of  only $\sim 2 \times 10^7 \rm M_{\odot}$, although with a large scatter between different simulations. In addition, the relative contributions of the three families of BH seed progenitors change and the total mass contributed by light BH progenitors is larger than that provided by medium-weight progenitors and comparable to that provided by heavy seeds.

\begin{figure}
     \hspace{-0.8 cm}

     \centering
     \includegraphics [width=9 cm]{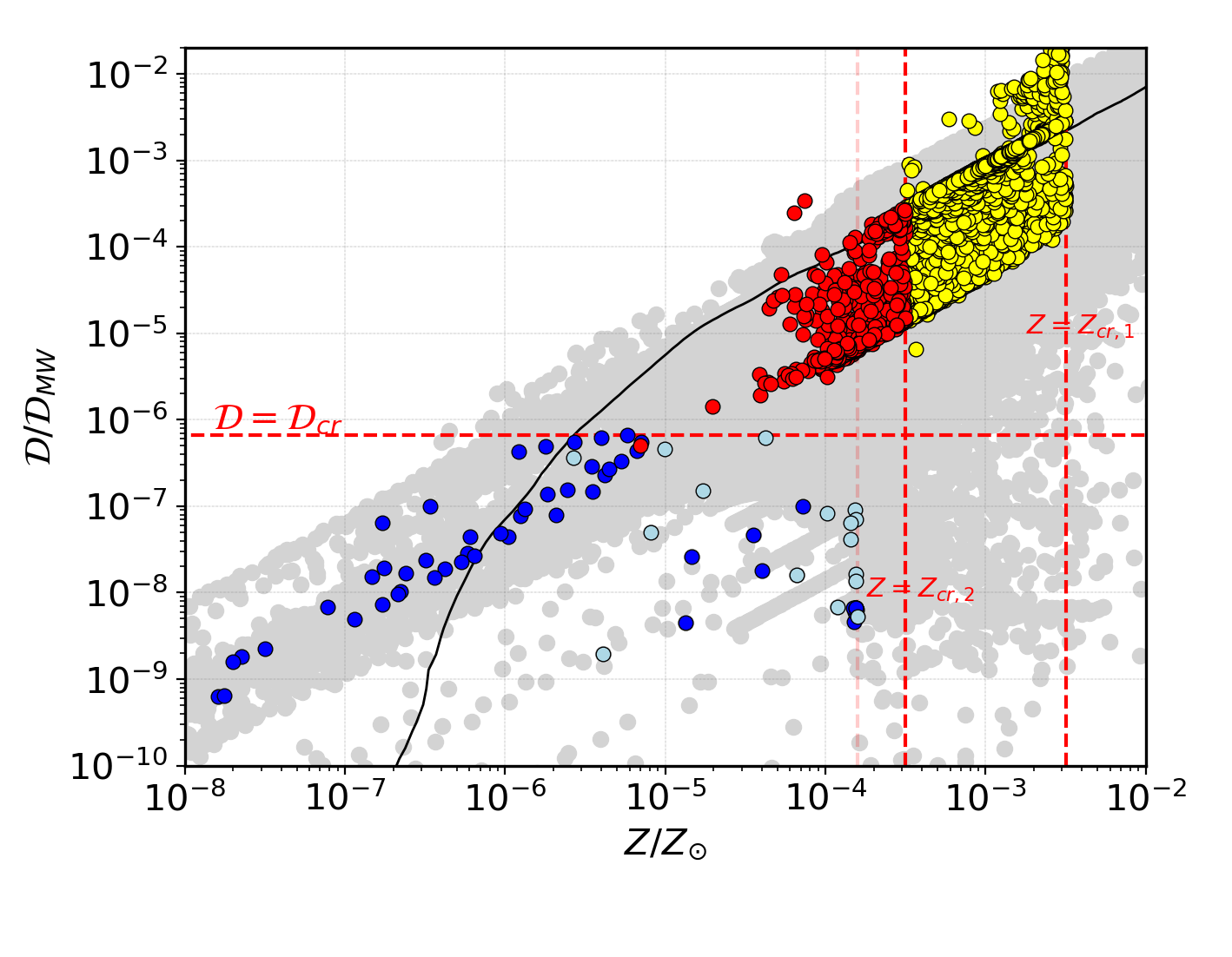}
     \caption{Same as Fig. \ref{fig:zvsdust} but for the set of parameters characterizing the SCA model and assuming $J_{\rm cr}=300$ (model SCA300). Red dashed lines are the critical dust-to-gas ratio and metallicity thresholds adopted in the SCA model (see Table \ref{tab:table2}). The thin vertical line shows the metallicity threshold adopted in the reference model (see Fig. \ref{fig:zvsdust} and Table \ref{tab:table2}).}  
     \label{fig:zvsdustSCA}
 \end{figure}
\begin{figure}
     \hspace{-0.8 cm}
     \centering
     \includegraphics [scale=0.186]{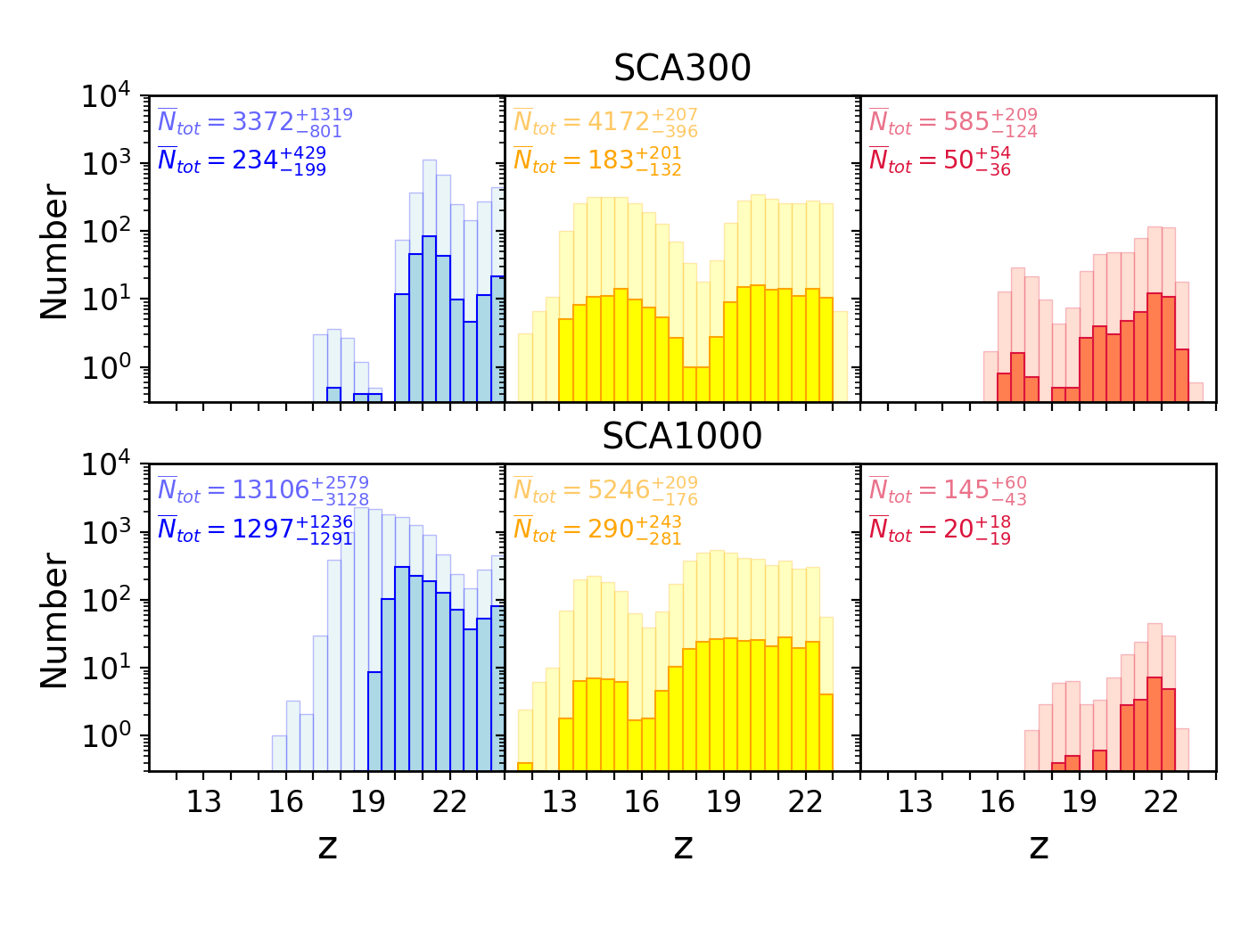}
     \caption{Same as Fig. \ref{fig:Rseeds} but for the set of parameters characterizing the SCA model and assuming $J_{\rm cr} = 300$ (model SCA300, top panels) and $J_{\rm cr} = 1000$  (model SCA1000, bottom panels).}
     \label{fig:Oseeds}
 \end{figure}
 \begin{figure}
      \hspace{-0.8 cm}
     \centering
     \includegraphics [scale=0.2]{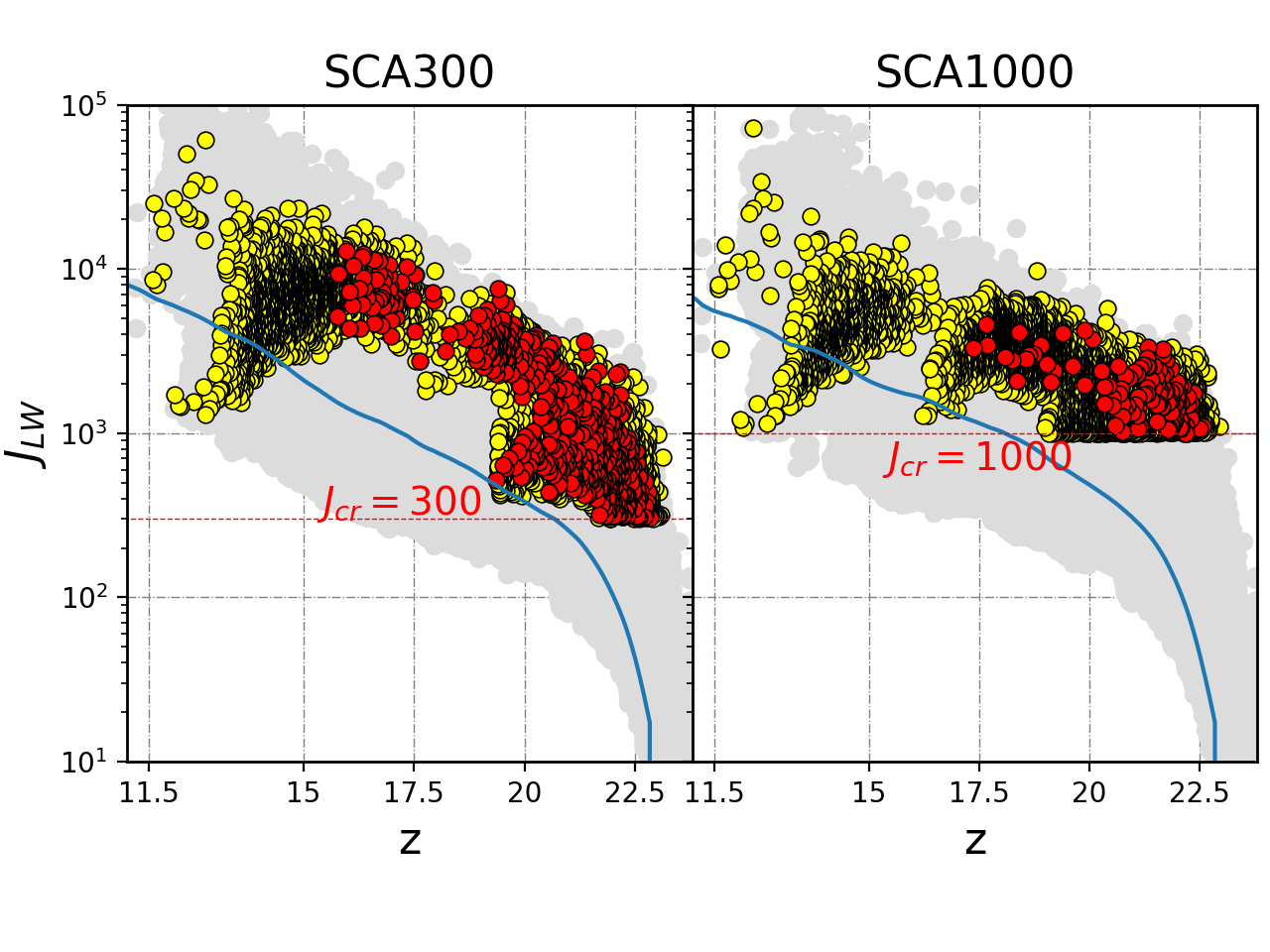}
     \caption{Same as Fig. \ref{fig:stdJ21} but for the set of parameters characterizing the SCA model and assuming $J_{\rm cr} = 300$ (model SCA300, left panel) and $J_{\rm cr} = 1000$  (model SCA1000, right panel).}
     \label{fig:OJ21}
 \end{figure}
 \begin{figure}
     \hspace{-0.8 cm}
     \centering
     \includegraphics [scale=0.36]{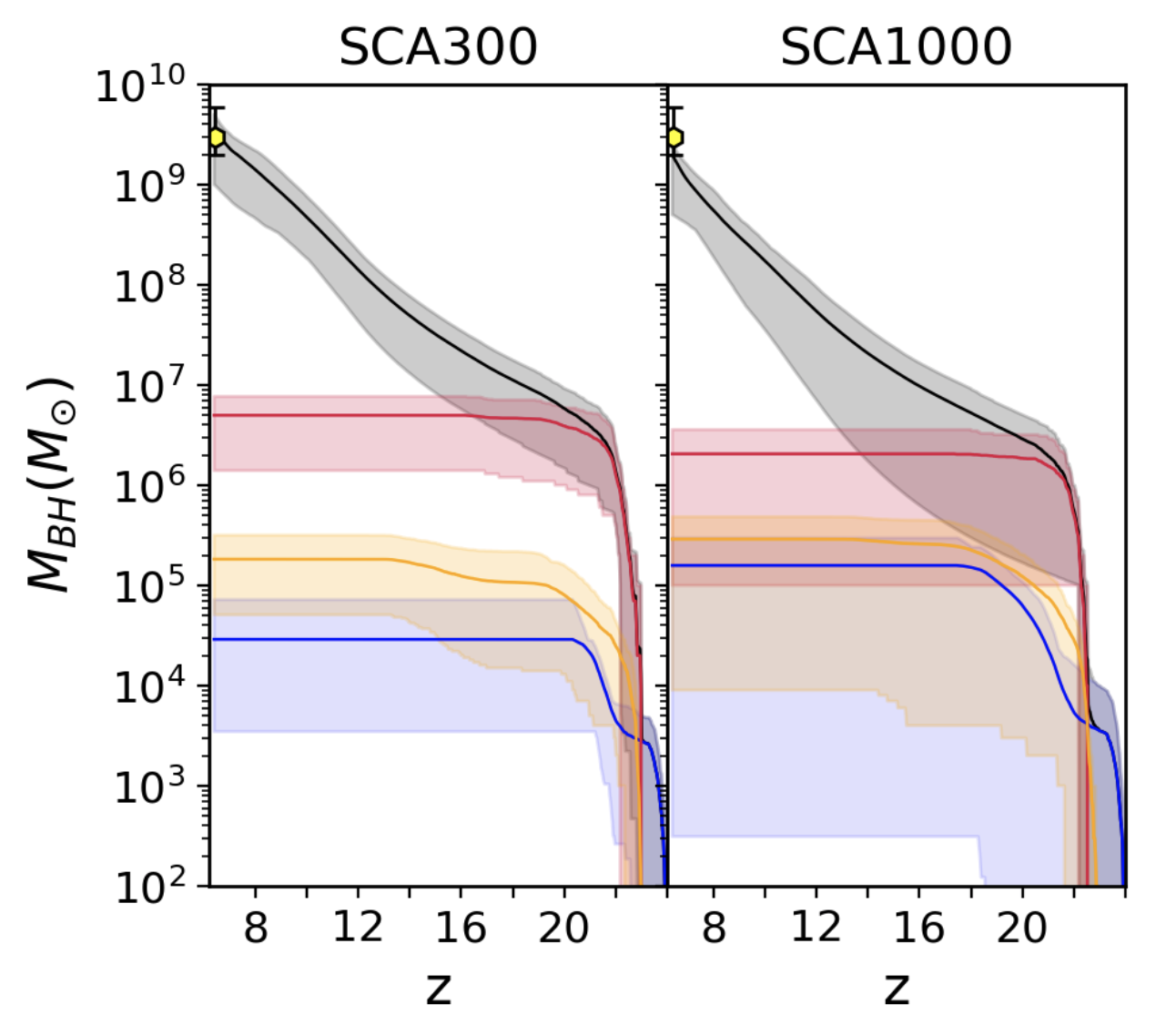}
     \caption{Same as Fig. \ref{fig:bhevoR} but for the set of parameters characterizing the SCA model and assuming $J_{\rm cr} = 300$ (model SCA300, left panel) and $J_{\rm cr} = 1000$  (model SCA1000, right panel).}
     \label{fig:Obhevo}
 \end{figure}

\subsection{Effects of super-competitive accretion}
\label{subsec:SCA}

As anticipated in section \ref{subsection:scamodel}, we have also investigated the effects of the recently
proposed SCA model \citep[][]{chon2020supermassive}. In this model, SMSs leading to 
medium-weight and heavy BH seeds may continue  to  form  at  higher  metallicities  than  assumed  in our  reference  model. In particular, we have run two additional sets of simulations, models SCA300 and SCA1000, where we have adopted
the initial conditions of the SCA model summarized in Table \ref{tab:table2} and two different values for $J_{\rm cr}$ ($300$ and $1000$, respectively). The results of these two additional models are illustrated in Figs.  \ref{fig:zvsdustSCA}, \ref{fig:Oseeds},
\ref{fig:OJ21},  and \ref{fig:Obhevo}.

We first compare the results of model SCA300 with the corresponding reference run, R300. 
 Note that, in order to ensure that the model reproduces the observed properties of our target quasar J1148 at $z=6.4$, we reduced the BH accretion efficiency, $\alpha_{\rm BH}$, from 150 (in the reference models) to 80 (in SCA models), as reported in Table \ref{tab:table1}.

Fig. \ref{fig:zvsdustSCA} shows non-pristine formation sites of the different seed populations in one single simulation of the SCA300 model, to be compared with the R300 results in Fig. \ref{fig:zvsdust}. The vertical and horizontal lines indicate the critical metallicity and dust-to-gas ratio thresholds adopted in the SCA model. As a consequence of the different conditions for their formation, both medium-weight and heavy seeds can form in a larger number of halos. Since in the SCA model dust-driven fragmentation does not prevent the formation of heavy seeds, some of the sites that were originally hosting middle-weight seeds in R300 can now lead to the formation of heavy seeds. 
This, however, does not reduce the number of medium-weight seeds as their formation can now occur in more metal enriched halos compared to the reference model.

Fig. \ref{fig:Oseeds} shows how the looser
constraints on metallicity and dust-to-gas ratio in their birth clouds lead to a significant increase in the 
number of medium-weight and heavy BH seeds, compared to the reference model ( by a factor of $\sim 9$ and $53$,
respectively). This increase is particularly dramatic for medium-weight BH seeds, that become more numerous than
light BH seeds in this model. The number of light BH seeds also increases ( by a factor of $\sim 1.8$),
but this is an indirect effect of a decrease in the LW background intensity (compare the left panels of Figures \ref{fig:stdJ21} and \ref{fig:OJ21}). Indeed, a large fraction of halos that were assumed to 
form Pop II stars in model R300 host the formation of medium-weight BH seeds in model SCA300, leading to a decrease
of the emitted LW radiation and hence of the global radiative feedback. This decrease favours the formation of
Pop III stars and light BH seeds compared to model R300. 
At the same time, the decrease in $J_{\rm LW}$ (and its fluctuations) does not compromise the formation 
of medium-weight and heavy BH seeds, that can occur in gas with higher metallicity in the SCA model. Figures \ref{fig:Oseeds} and \ref{fig:OJ21} show that in model SCA300 the formation of medium-weight and heavy seeds extends to lower redshift and is hosted in halos illuminated by lower $J_{\rm LW}$ compared to model R300. 

Similar considerations apply if we restrict the analysis to BH seed progenitors: the darker coloured histograms in
Fig. \ref{fig:Oseeds} show that they follow the same redshift distribution of their parent BH populations.
Compared to model R300, the number of medium-weight and heavy BH seed progenitors formed in model SCA300 
increases by a factor $\sim 3.3$ and $16.6$, respectively.

The results change when the value of $J_{\rm cr}$ is increased to 1000, as in model SCA1000:
compared to model SCA300, the total number of medium-weight seeds mildly increases, heavy seeds are a  factor of $\sim 4$ less numerous
and the number of light seeds increases by a factor of $\sim 3.8$. Hence, in this case, the milder radiative
feedback favours Pop III star formation, and the associated metal enrichment disfavours the formation of
heavy BH seeds.  The BH seed progenitor populations are also affected, with heavy BH seeds being a factor $\sim 2.5$ less numerous than in model SCA300, while medium-weight and light BH seed progenitors increasing by  $\sim 1.6$ and $\sim 5.5$, respectively.

It is interesting to see how these differences affect the nuclear BH growth. 
A comparison between  Figs. \ref{fig:bhevoR} and \ref{fig:Obhevo} shows that the more favourable conditions to heavy and medium-weight seeds formation in SCA models
lead to the successful growth of a $\sim 10^9 \rm M_{\odot}$ SMBH at $z \sim 6$, independently of the adopted value of $J_{\rm cr}$. Despite the total BH mass contributed
by medium-weight seeds is one order of magnitude higher than in the R300 and R1000 models, the results of SCA models confirm that the dominant mass contribution is provided by the less numerous but more massive heavy seeds.

\begin{figure}


    \centering
    
    \includegraphics [scale=0.19]{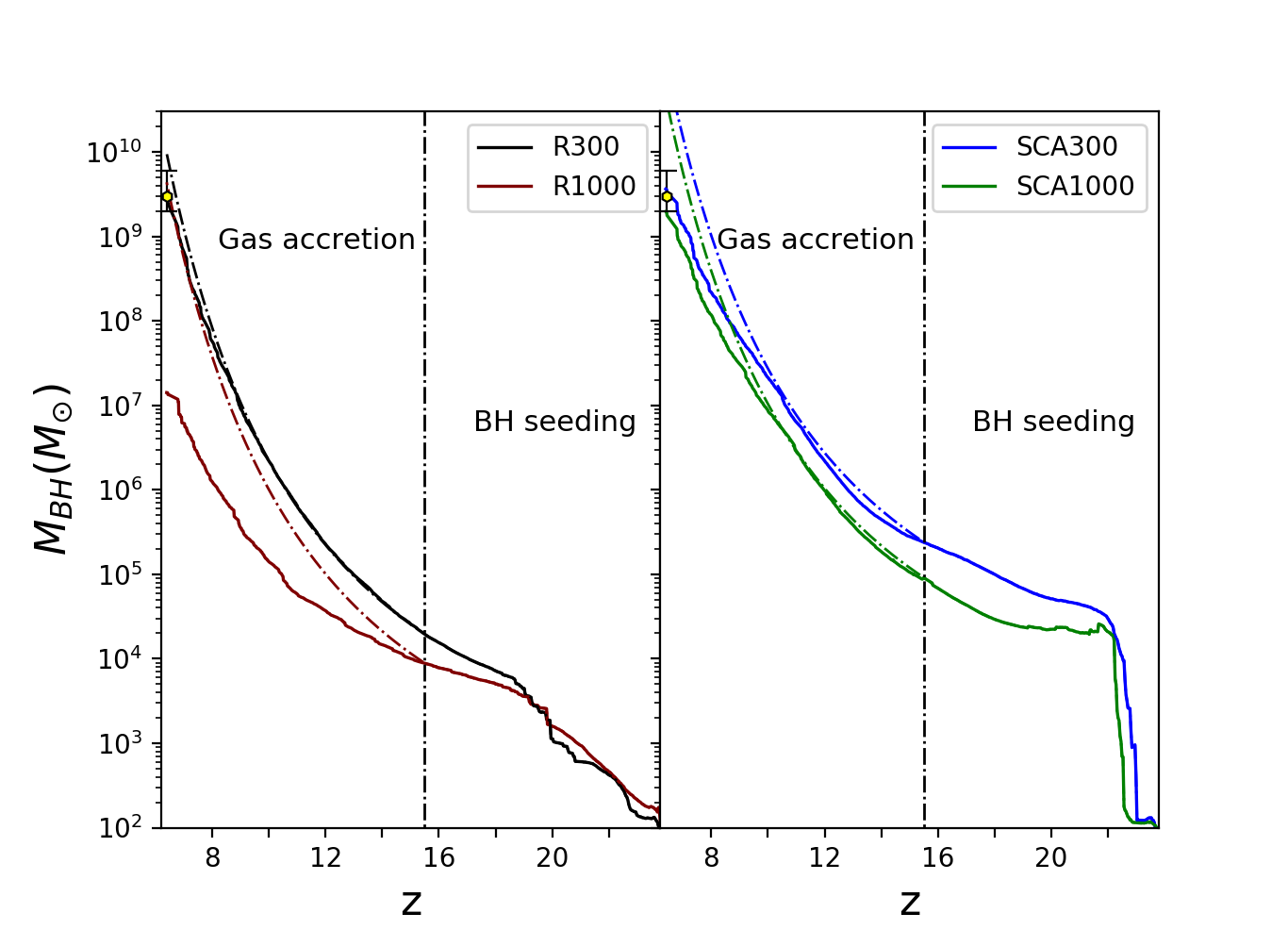}
    \caption{ Left panel: evolution of the mean BH mass as a function of redshift in models R300 (black), R1000 (red). Each line is the average value among ten simulations. The vertical line indicates the epoch when seed formation terminates at $z \sim 15.5$. Below this redshift, nuclear BHs mostly grow by gas accretion. We compare the mean BH mass growth predicted by \texttt{GQd} models (solid lines) with the growth histories of BHs with a mass equal to the mean nuclear BH mass at $z \sim 15.5$ if these were assumed to grow uninterruptedly at the Eddington rate (dashed lines). Right panel: same as in the left panel but for models SCA300 (blue) and SCA1000 (green).}
    \label{fig:meanBH}
    
\end{figure}

\begin{figure}

     \hspace{-0.8 cm}

    \centering

    \includegraphics [scale=0.2]{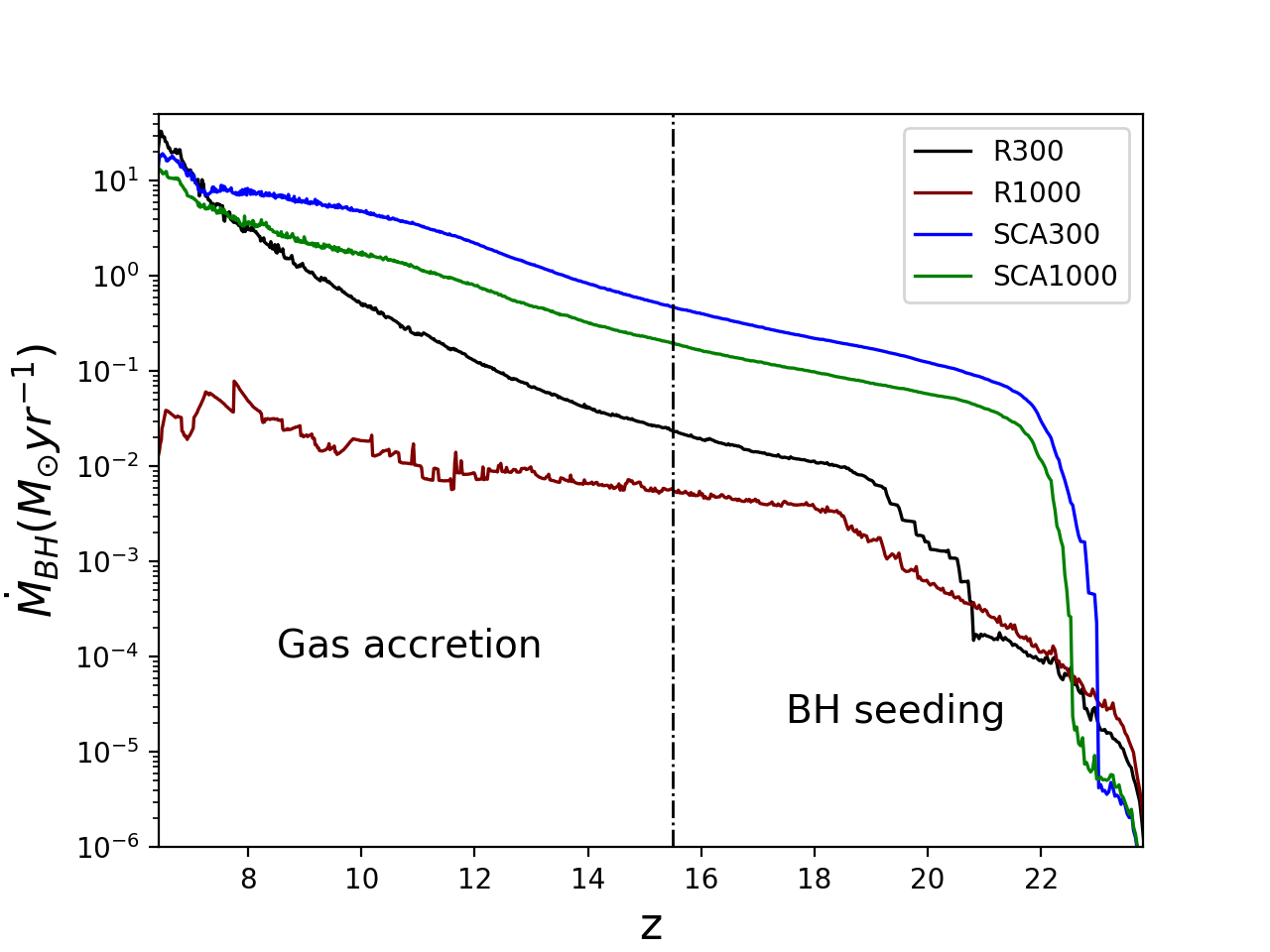}
 
    \caption{ Evolution of the mean BH accretion rate as a function of redshift in models R300 (black),  R1000 (red), SCA300 (blue), and SCA1000 (green). Each line is the average value among ten simulations. The vertical line indicates the epoch when seed formation terminates, at $z \sim 15.5$.}
    \label{fig:meanBHacc}

 \end{figure}

\subsection{A composite picture: quality not quantity}

The analysis presented above highlights how sensitive the genealogy of $z \sim 6$ SMBHs is to 
BH seeds birth conditions. We find that the roots of the family tree are populated by
light, medium-weight, and heavy BH ancestors, but their relative number and frequency is very
sensitive to the combination of metallicity, dust-to-gas mass ratio, and illuminating 
LW flux that are assumed to provide the right conditions for their formation. 

At early times, when $z \ge 16$, the evolution is dominated by BH seeding and the family
tree that emerges can be densely populated, with a total mass in nuclear BHs that can be larger than $\sim 10^6 \rm M_{\odot}$. However, we also find that a successful SMBH 
growth history relies on the quality of its BH seeds progenitors rather than on their quantity.
Indeed, more than 99\% of the final SMBH mass must come from gas accretion, that drives the
evolution at later times, when $z < 16$. 

In our Eddington-limited accretion scenario, BHs grow at the
Bondi rate that is very sensitive to BH mass (see Eq. \ref{eq:BHL}). Hence, for the same total mass of
BH seeds formed, family trees where the BH seeds mass distribution is more top heavy and the
mean BH mass is larger are better equipped to grow to a $> 10^9 \rm M_{\odot}$ SMBH.

This is illustrated in Fig. \ref{fig:meanBH}, where we show the evolution of the mean BH mass 
as a function of redshift for the reference and SCA models, adopting two different values of 
the critical LW flux ($J_{\rm cr} = 300$ and $1000$).

As expected, at early times, during the BH seeding phase, the mean BH mass is strongly affected by the 
relative weight of different BH populations. 

At the end of the BH seeding phase, the different seed BH mass distributions lead to mean BH masses that
range between $\sim 10^4 \rm M_{\odot}$ for model R1000 to $3 \times 10^5 \rm M_{\odot}$ for model SCA300. 
During the early phase of the accretion-dominated evolution, a successful SMBH growth history requires rapid 
mass growth, close to the Eddington rate. To emphasize this point, for each model variant, the dashed lines 
show what would be the evolution if a single BH with a mass equal to the mean BH mass at $z \sim 16$ were 
assumed to evolve by growing uninterruptedly at the Eddington rate.
The comparison between the solid and dashed lines, for each model, shows that {\rm with exception of model
R1000, the mean BH mass initially grows very close or at the Eddington rate, and becomes sub-Eddington during 
the late phase of the evolution, at $z < 10 - 12$.} 

 Fig. \ref{fig:meanBHacc} shows the mean BH accretion rate as a function of redshift for the four models. 
During the BH seeding epoch, at $z \geq 15.5$, the accretion rates are affected by the evolving mass
distribution of BH seeds progenitors. At all but the lowest redshifts, SCA models are characterized by the largest
mean BH accretion rates, reflecting their larger mean BH mass. In model R300, the accretion rate
increases more steeply than in SCA models, as a result of the larger gas mass in progenitor halos, that are 
less affected by AGN feedback. Finally, in model R1000 the mean accretion rate never exceeds $0.1 \, \rm M_\odot/yr$
and shows a larger stochasticity, reflecting the smaller mean BH masses and the lower number of BH seeds in this
model.

\begin{figure*}
     \centering
     \includegraphics [scale=0.2]{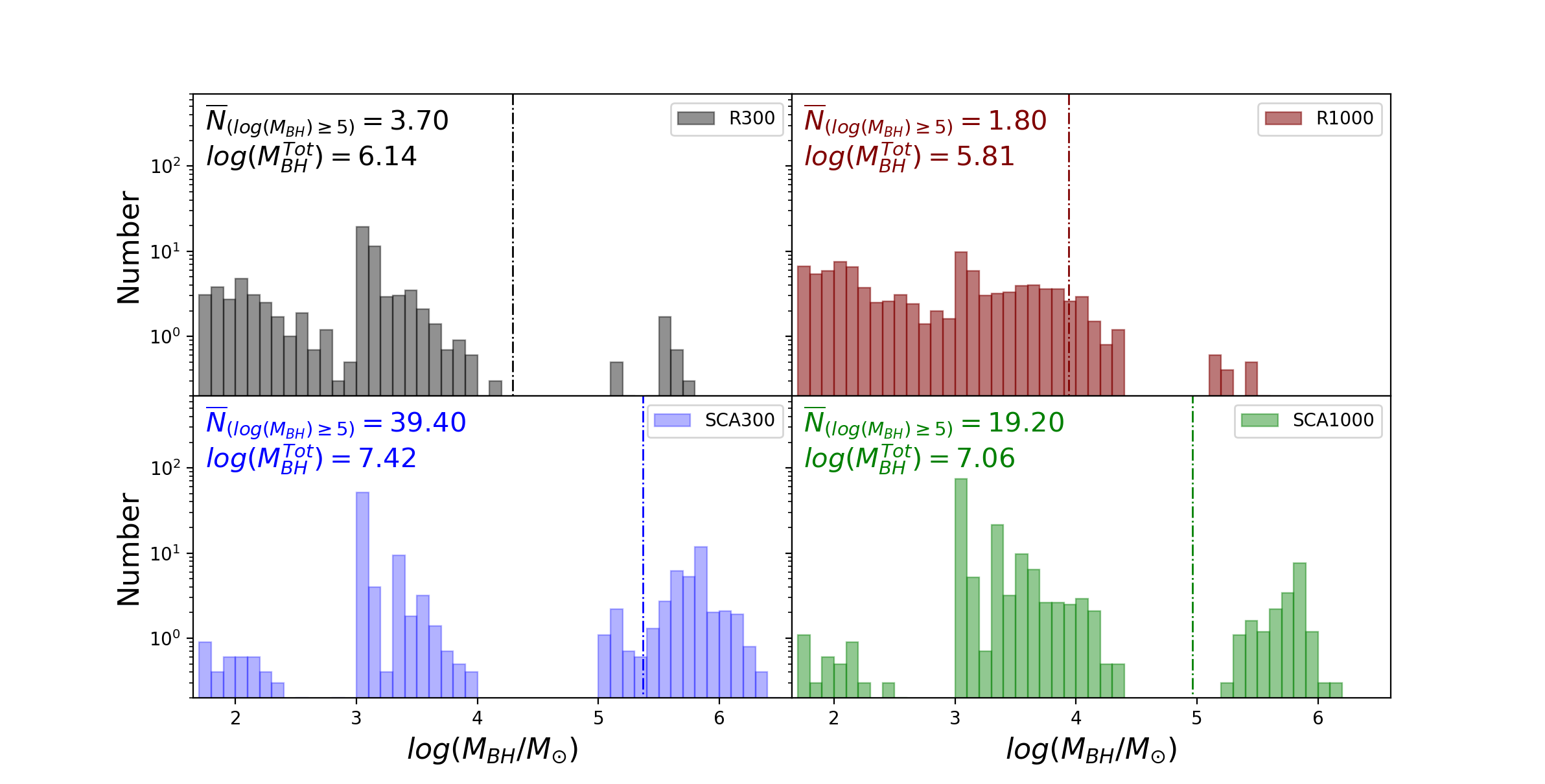}
     \caption{The mean BH mass function at $z=15.5$ averaged over 10 simulations of models R300 (top-left), R1000 (top right), SCA300 (bottom-left), and SCA1000 (bottom right). In each panel, the vertical line indicates the mean BH mass, whereas the numbers refer to the total mass and number of BHs with $M_{\rm BH} \geq 10^5 \rm M_{\odot}$.}
     \label{fig:BHmassfunc}
 \end{figure*}

This result can be understood by looking at the
mass distributions of BH seed progenitors that characterize the four models at the end of the BH seeding epoch. Fig. \ref{fig:BHmassfunc} shows the average distribution over 10 simulations at $z = 15.5$. 
At the end of the BH seeding epoch, some BH progenitors have merged to form higher mass systems (i.e. the BHs in range $10^4 <\rm M_{\rm BH}/\rm M_{\odot}<10^5$ or with $\rm M_{\rm BH}\geq 10^5 \rm M_{\odot}$). The vertical
lines in each panel indicate the mean BH mass, whereas the numbers refer to the total mass and number of BHs
with $\rm M_{\rm BH}\geq 10^5 \rm M_{\odot}$. These are the systems that drive the mass growth in the accretion-dominated
phase at $z < 15.5$. In models R300 and R1000, the number and total mass of BHs in this 
mass range is a factor of $\sim 10$ smaller than in models SCA300 and SCA1000, reflecting the difference
in heavy BH seeds birth rate among these models. While in model R300 the enhanced accretion efficiency can counterbalance this large discrepancy, in R1000 the subsequent growth rate is too slow and leads to a final nuclear BH mass $\sim 2 \times 10^7 \rm M_{\odot}$ at $z = 6.4$, as shown by right panels of
Figs. \ref{fig:bhevoR}. To appreciate the large variety of possible growth histories
among the different simulations, for each of the four models we report in the Appendix the 
BH mass evolution in each of the 10 runs, highlighting the contribution of light, medium-weight, and heavy BH seed progenitors.
 
\section{Discussion}
\label{section:discussion}

Investigating the origin and growth history of $z \ge 6$ SMBHs is very challenging. While a number
of studies have been devoted to explore BH seed formation \citep[][]{valiante2017} and their subsequent
evolution through mergers and gas accretion \citep[][]{inayoshi2020}, an ab-initio, self-consistent 
theory is still lacking. 

Large-scale cosmological simulations such as MassiveBlack \citep[][]{dimatteo2012}, Illustris \citep[][]{vogelsberger2014}, EAGLE \citep[][]{schaye2015}, MassiveBlack-II \citep[][]{Khandai2015} and 
BlueTides \citep[][]{feng2016} have been able to achieve good agreement with observations by 
implementing subgrid prescriptions for BH seed formation, which generally adopt a fixed 
BH seed mass that is planted in halos above a given mass threshold.
In these approaches, both the BH seed mass and the minimum halo mass are free parameters
of the simulation. More recently, attempts have been made to include in 
smaller-scale cosmological simulations or in zoom-in/constrained simulations more physically 
motivated prescriptions for BH seeding \citep[][]{bellovary2011, habouzit2017, tremmel2017, huang2020}. 
Yet, the exploration of different seed physical parameters or their growth history is still 
too computationally expensive, despite attempts being made to avoid this limitation through
post-processing methods \citep[][]{degraf2020}.

Here we have followed a complementary approach and have used semi-analytical models to run independent
evolutionary histories of a $10^{13} \rm M_{\odot}$ dark matter halo at $z = 6.4$, following the evolution
of its baryonic content since the onset of star and black hole formation at $z = 30$. Building on a
number of previous studies \citep[][]{valiante2011origin, valiante2014high, valiante2016first, pezzulli2016, pezzulli2017sustainable, valiante2018statistics}, here we have explored, for the first time, the incidence and relevance of three different BH seed populations for the early formation of SMBH. One important
aspect of our investigation is that we do not restrict the analysis to a single BH seed population. 
Rather, we populate the roots of SMBHs family trees with light, medium-weight, and heavy BH seeds 
depending on their specific birth conditions.  In particular, here we have improved our previous
model \citep[][]{valiante2016first} by ({\it i}) introducing the possibility to form 
medium-weight BH seeds and by ({\it ii}) implementing a new treatment of radiative, and chemical feedback that accounts for spatial fluctuations in the metal (dust) distribution and in the LW radiation field.

Despite these new features, our main conclusions agree with previous findings, demonstrating, for the first time, that the formation of medium-weight BH seeds does not qualitatively change the growth history of the first SMBHs. Indeed, similarly to \citet{valiante2016first, valiante2018statistics}, we find that, if BHs are allowed to grow without exceeding their Eddington rate, the formation of the first SMBHs at $z > 6$
relies on a small number of heavy BH seed progenitors. 
At the same time, we also find that medium-weight BHs are very common among the ancestors of 
$z = 6.4$ SMBHs, particularly in the SCA scenario proposed by \citet{chon2020supermassive}.
This is in broad agreement with the pioneering work of \citet{devecchi2009formation, devecchi2012},
who also investigated when and where BHs could form as a result of runaway collisions of massive stars in
dense nuclear star clusters. Here we have targeted our study to investigate the formation history of a
prototypical quasar at $z = 6.4$ instead of sampling a more representative cosmological volume. 
As a result, we find that despite medium-weight BHs always outnumber heavy BH seeds, 
their contribution to SMBH growth at $z > 6.4$ is always subdominant with respect to that of heavy BH seeds. 

One important caveat of our analysis concerns the assumed BH seed masses. While light BH seed masses are 
randomly sampled from an underlying Pop III stellar mass function and are therefore characterized by a
mass distribution extending from a few tens to maximum 300 $\rm M_{\odot}$, we have adopted a constant
mass of $10^3 \, \rm M_{\odot}$ and $10^5 \, \rm M_{\odot}$ for medium-weight and heavy BH seeds, respectively. This is clearly an oversimplification, as it is expected that, depending on the physical conditions at place, 
medium-weight seeds may form with a range of masses extending from $\sim 300$ to $\sim 3000 \, \rm M_{\odot}$ \citep[][]{devecchi2012} or $\sim 400$ to $\sim 1900 \, \rm M_{\odot}$ \citep[][]{sakurai2017formation}, or $\sim 100$
to $\sim 5000 \, \rm M_{\odot}$ \citep[][]{reinoso2020}. Similarly, heavy BH seeds are also expected to span
a mass range that varies between $\sim 10^3$ to $[3 - 5] \, 10^5 \rm M_{\odot}$ \citep[][]{lodato2007}, or $\sim 5\times 10^5$ and $\sim 2\times 10^6 \, \rm M_{\odot}$ \citep[][]{ferrara2014}, or even to form in binary systems with masses of $10^3$ - $10^5 \, \rm M_{\odot}$ \citep[][]{chon2018}. In addition, these figures may be significantly altered in the SCA scenario \citep[][]{chon2020supermassive}. Hence, the mass ranges of the three BH seed populations are not yet firmly assessed and may partly overlap, forming a continuum distribution from few tens to
few $10^5 \, \rm M_\odot$. 

 Finally, it is worth mentioning that the formation of direct collapse BHs may be driven/triggered also by the combined effect of sub-critical LW radiation and dynamical heating, both contributing in suppressing fragmentation/star formation in rapidly growing pre-galactic halos, so that, in overdense regions, heavy seeds could be more common (larger number density) than previously expected \citep[e.g.][]{wise2019formation}. 

While a more refined description of the seed mass and its dependence on the birth conditions is deferred to a future study, here we anticipate that this may impact our results by
changing the relative contribution of the three BH seeds populations to the mass growth of 
the final SMBH. More indirectly, it may also modify the baryonic evolution of their host galaxies
and hence the relative frequency of medium-weight and heavy BH seeds that form
at later times. Yet, our results suggest that, for each SMBH in place at $z = 6.4$, a large number of BH seeds form 
at $z > 15 - 16$ (see Figs. \ref{fig:Rseeds} and \ref{fig:Oseeds}), only a small fraction of
which are the progenitors of the final SMBH.
 Another caveat of our model is that we do not consider baryonic streaming motion (BSM), that may suppress star formation in less massive halos (typically $T_{\rm vir}<10^4 K$), preventing gas collapse \citep[][]{schauer2017, inayoshi2018massive, hirano2018formation}. This limitation does not have a direct impact on the formation of heavy and medium-weight BH seeds, which form in larger, atomic cooling halos. However, the lack of star formation in mini-halos may have an indirect effect because (i) the gas will remain chemically pristine until atomic-cooling halos form and (ii) the suppression of H$_2$ cooling will be possible only by nearby star forming atomic cooling halos, in the so-called "syncronized-halo" model \citep{visbal2014}. While it is not straightforward to assess the net effect of these two processes in our model, \citet{schauer2017} argue that BSM may provide favourable conditions
for the formation of heavy BH seeds. The minimum dark matter halo mass where gas cooling is suppressed by BSM increases ($T_{\rm vir} \sim 1-2 \cdot 10^4\,$K) if the halo experiences a violent merger episode \citep{inayoshi2018massive}. However these events are expected to be rare, affecting
less than $1\%$ of the halos (see Fig. 2 in \citealt{inayoshi2018massive}).

\section{Summary and Conclusions}
\label{section:summary}

We have investigated the role of three BH seed populations in the formation history of $z > 6$ SMBHs.
Building on previous studies, we have further extended our semi-analytical data-constrained model, \texttt{GQd},
by including a statistical description of the spatial fluctuations in metal and dust enrichment and 
in the intensity of the LW radiation field. For the first time, we have investigated the birth rate
of light, medium-weight, and heavy BH seeds by selecting their formation sites according
to physically motivated conditions on the metal and dust content as well as on the intensity of the
illuminating LW radiation field (model R300). We find that:

\begin{itemize}
    \item Inhomogeneous metal and dust enrichment and fluctuations in the LW radiation favour the
    formation of seeds and extends their birth epoch down to $z \sim 16$, when the filling
    factor of enriched regions becomes $Q \sim 1$;
    \item Light seeds are the first to form in pristine star-forming regions, and their formation
    is suppressed at $z \sim 20$ by radiative feedback that illuminates the metal-poor gas above
    the critical value, preventing H$_2$-cooling. On average, $\sim 1831$ light seeds form in our
    reference model, but less than $\sim 13$ per cent of these are true progenitors of the final SMBH;
    \item Medium-weight and heavy BH seeds form at $16 \leq z \leq 23.5$ when atomic cooling halos
    start to assemble. Medium-weight seeds form in metal-poor star-forming regions illuminated by
    a LW radiation $J_{\rm cr} < J_{\rm LW} < 100 \, J_{\rm cr}$ from neighbouring systems,
    while heavy seeds mostly form in pristine regions when the illuminating LW intensity reaches
    values $10 \, J_{\rm cr} < J_{\rm LW} < 100 \,J_{\rm cr}$. On average, $\sim 467$ medium-weight and
    $\sim 11$ heavy seeds form in the reference model, $\sim 12 - 27 \%$ per cent of which are 
    SMBH progenitors;
    \item Despite their smaller number, heavy BH seed progenitors provide the largest contribution
    to the SMBH mass growth, with a cumulative mass of $\sim 3 \times 10^5 \rm M_{\odot}$ and triggering efficient
    gas accretion, that drives the mass growth at $z < 16$ and leads to the formation of a $\sim 3 \times 10^9 \rm M_{\odot}$ SMBH at $z = 6.4$. For most of the time, the average gas accretion rate is very close to Eddington.
\end{itemize}

We have explored the sensitivity of the above results to variations of the adopted critical value
of the LW flux by running a set of simulations with $J_{\rm cr}=1000$ (R1000). We find that:
\begin{itemize}
\item A larger $J_{\rm cr}$ has the effect of reducing the strength of radiative feedback,
favouring Pop III star formation and increasing the number of light seeds ( by a factor $\sim 5$).
However, this also causes a more efficient metal and dust enrichment, which limits the formation
of medium-weight and heavy black hole seeds, decreasing their number by a factor $\sim 3.3-5.5$. A similar
decrease ( by a factor of $\sim 2.2-3$) is also found in the number of medium-weight and heavy seed progenitor.
This has a large impact on the formation of the SMBH, which on average reaches a mass of $\sim 2\times 10^7 \, \rm M_{\odot}$ by $z = 6.4$, with a large dispersion of values among different evolutionary histories.
\end{itemize}

Finally, we have explored the possibility to form SMSs - leading to medium-weight and heavy BH seeds - at higher metallicity than assumed in our reference model, thanks to super-competitive accretion
\citep[][]{chon2020supermassive}. We find that:

\begin{itemize}
    \item The looser constraints on metallicity and dust-to-gas ratio in the birth clouds predicted 
    by this model (SCA300) lead to a significant increase in the total number of medium-weight (by a factor $\sim 9$) and heavy (by a factor $\sim 50$) BH seeds; the number of light seeds also increases (by a factor $\sim 2$) due to the lower intensity of the LW background and the milder radiative feedback induced by a more significant contribution to the LW emissivity by accreting BHs, that are less efficient sources compared to Pop II stars. Heavy seeds
    dominate the mass growth, leading to the formation of a $\sim 3 \times 10^9 \rm M_{\odot}$ SMBH at $z = 6.4$, even when the critical intensity of the LW background is increased to $J_{\rm cr}=1000$ (SCA1000).
\end{itemize}

In the Eddington-limited accretion scenario that we have explored, we find that a successful SMBH growth history relies on the quality of its BH progenitors, rather than on their quantity: family trees whose roots are seeded by a sufficient number of heavy BH progenitors are better suited to grow $> 10^9 \, \rm M_{\odot}$ SMBHs by $z = 6.4$. 

 Our study suggests that the genealogy of $z \sim 6$ SMBHs is characterized by a rich variety of BH progenitors, which represent only a small fraction ($< 10 - 20 \%$) of all the BHs that seed galaxies at $z > 15.5$. While their mass distribution depends on the physical conditions (metallicity, dust-to-gas ratio, illuminating LW radiation) at their birth, these properties need to be explored in the attempt of making a census of the BH population at high-$z$. The present study is a first attempt
to describe the richness of the BH landscape at cosmic dawn that may be explored by future electromagnetic and gravitational wave facilities.

\section*{Acknowledgements}
We thank the Referee for her/his careful reading of the manuscript and insightful comments.\\
We acknowledge support from the Amaldi Research Center funded by the MIUR program \lq \lq Dipartimento di
Eccellenza \rq \rq (CUP:B81I18001170001), from the INFN TEONGRAV specific initiative, and the networking support by the COST Action CA16104.
\\ This work is partially supported by the National Science Foundation of China (11721303, 11991052, 11950410493), the National Key R\&D Program of China (2016YFA0400702).
\\ KO and SC acknowledge financial support by the Grants-in-Aid for Basic Research by the Ministry of Education, Science and Culture of Japan (SC:19J00324,KO:25287040, 17H01102, 17H02869).
\\F.S. thanks the University of Z{\"u}rich for the kind support and hospitality 
and acknowledges the precious exchange of opinions and helpful advice of participants to the S$\tilde{a}$o Paulo School of advanced science on FIRST LIGHT and XXXI Canary Islands Winter School of Astrophysics.

\section*{Data Availability}

The simulated data underlying this article will be shared on reasonable request to the corresponding author.
The code \texttt{GQd} is not publicly available. The adopted approach and further possible improvements can be discussed with the corresponding author and the other code developers/users within the working group.




\bibliographystyle{mnras}



\appendix

\section{BH mass growth}
\begin{figure*}
     \centering
     \includegraphics [scale=0.25]{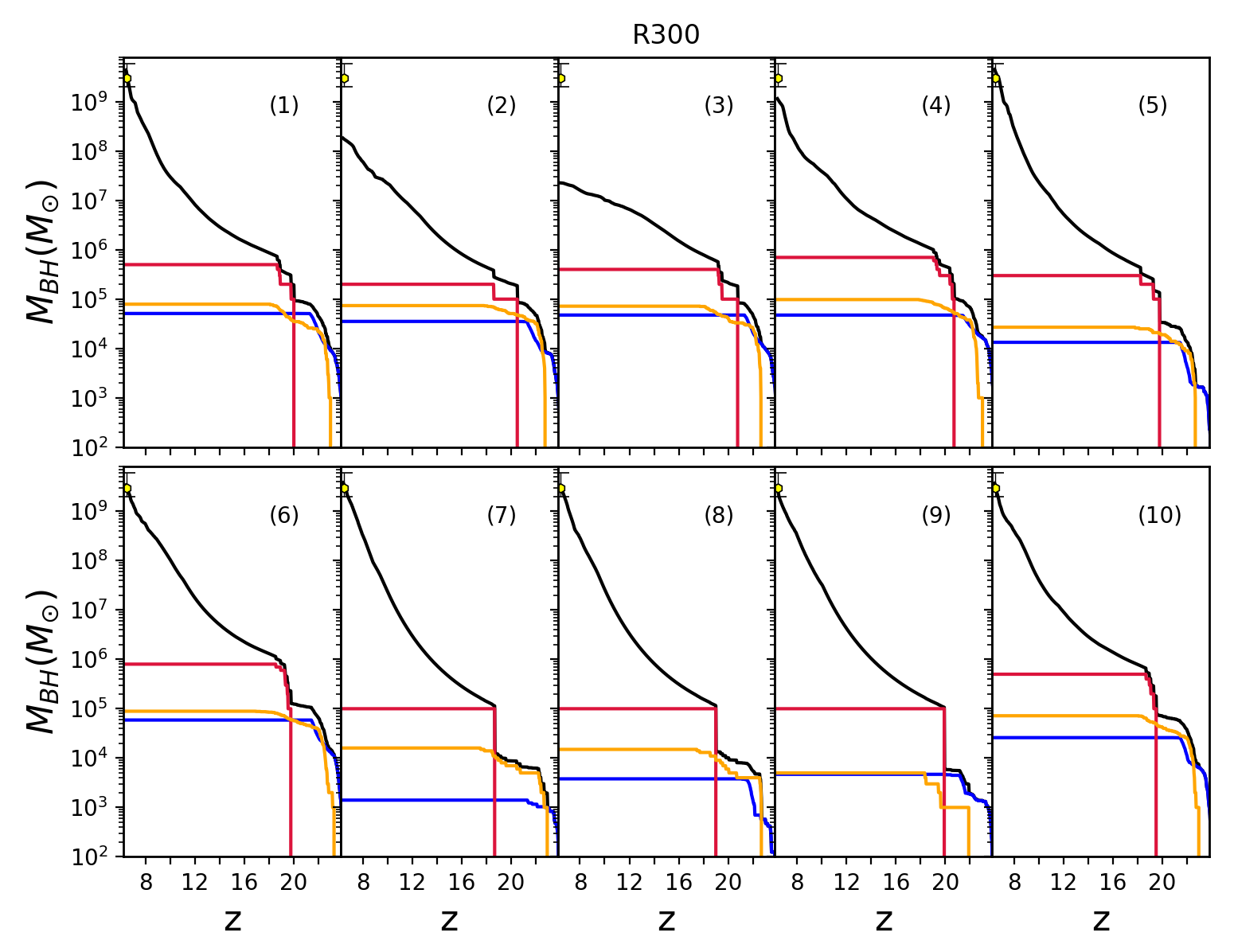}
     \caption{Evolution of the total mass of nuclear black holes as a function of redshift (black line) with the separate contributions of light (blue line), medium-weight (yellow line), and heavy BH seed (red line) progenitors. Each panel shows individual results over 10 simulations for R300 model.}
     \label{fig:10bhR300}
 \end{figure*}

\begin{figure*}
     \centering
     \includegraphics [scale=0.25]{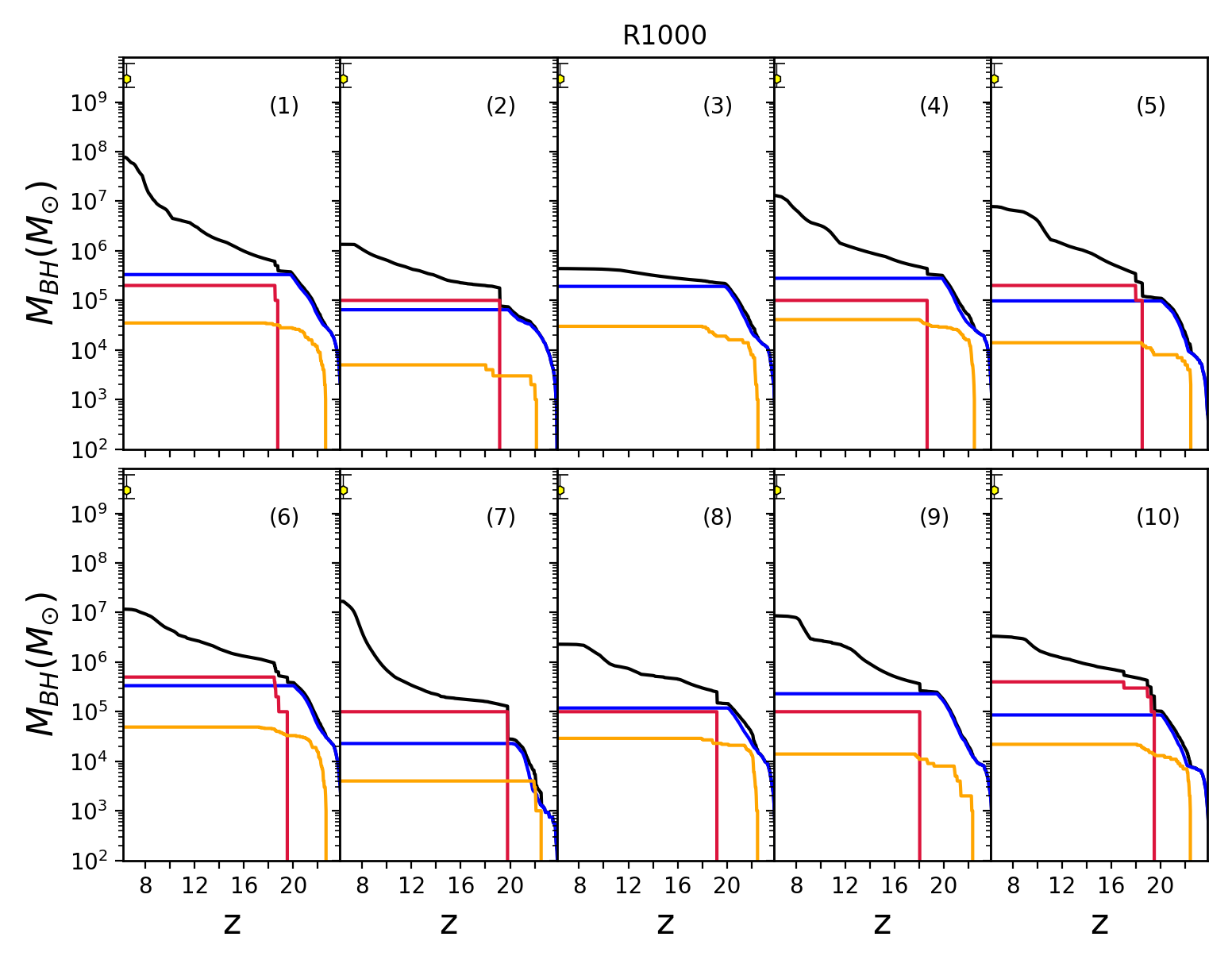}
     \caption{Same as Fig. \ref{fig:10bhR300} but for model R1000.}
     \label{fig:10bhR1000}
 \end{figure*}

\begin{figure*}
     \centering
     \includegraphics [scale=0.25]{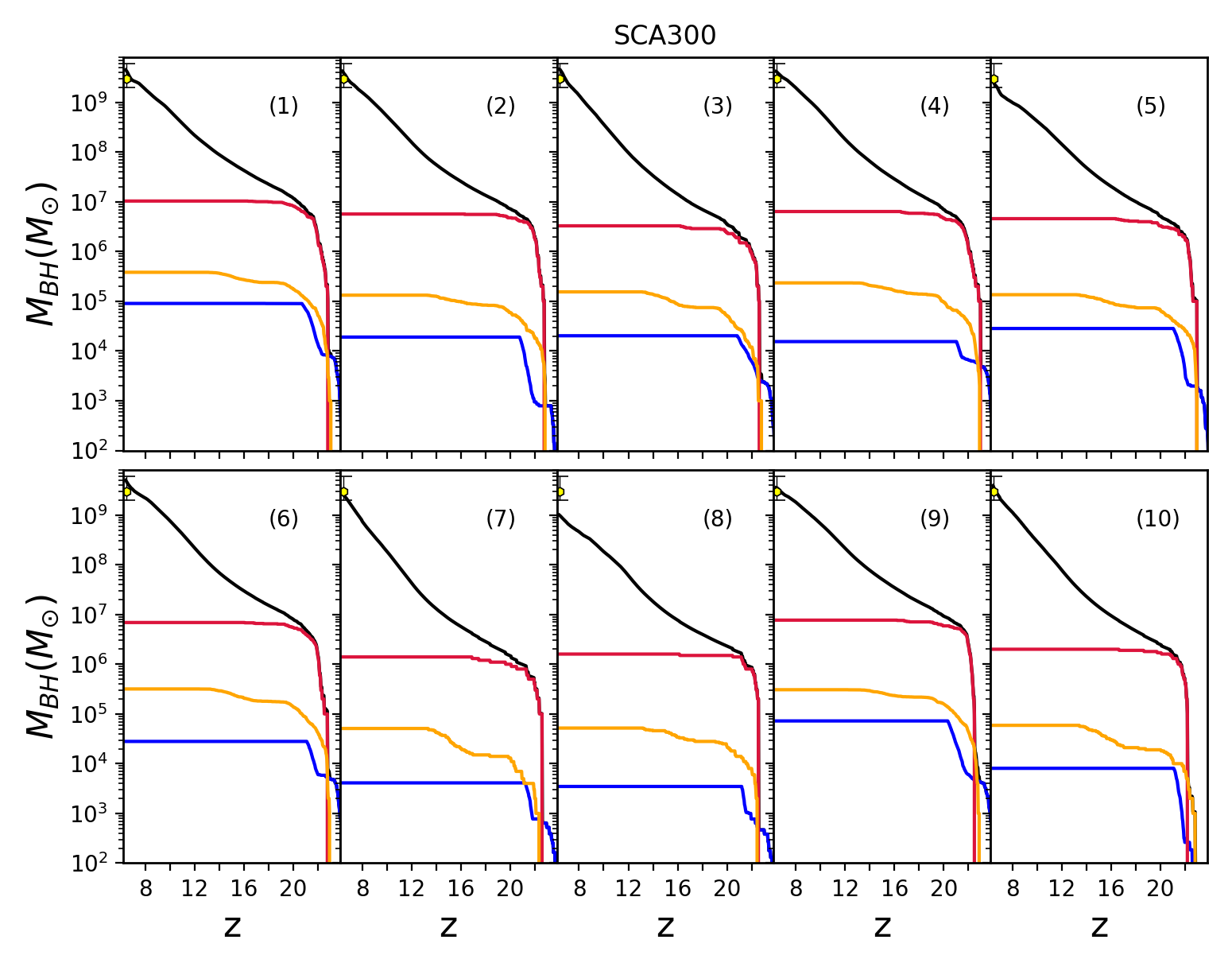}
     \caption{
     Same as Fig. \ref{fig:10bhR300} but for model SCA300.}
     \label{fig:10bhSCA300}
 \end{figure*}

\begin{figure*}
    
     \centering
     \includegraphics [scale=0.25]{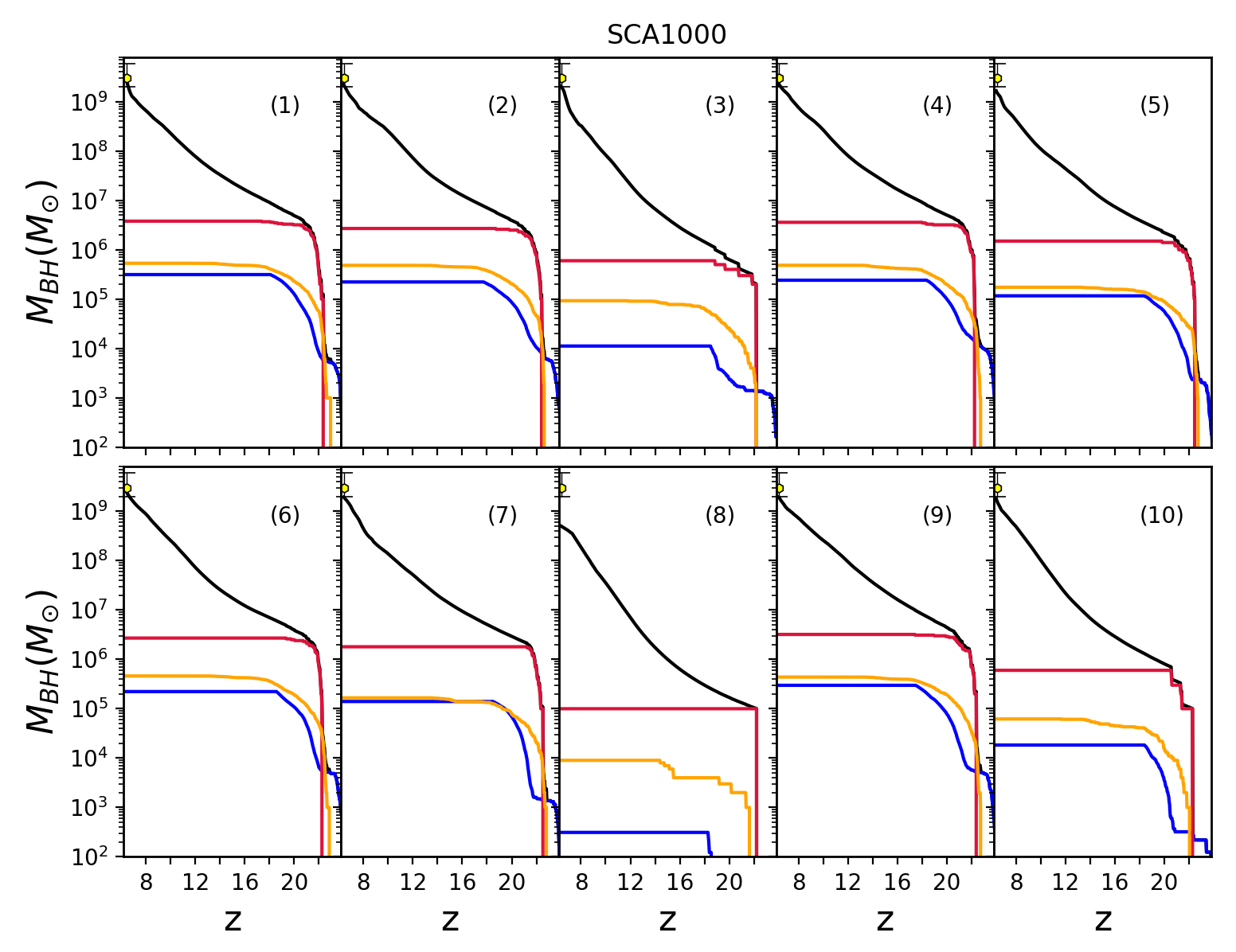}
     \caption{Same as Fig. \ref{fig:10bhR300} but for model SCA1000.}
     \label{fig:10bhSCA1000}
 \end{figure*}

In sections \ref{subsec:Nuclearbh} and \ref{subsec:SCA}, we have discussed the evolution of the total mass of nuclear BHs as a function of redshift, highlighting the contribution of light, medium-weight, and heavy seeds in the R300, R1000, SCA300, and SCA1000 models. Figures \ref{fig:bhevoR} and \ref{fig:Obhevo} illustrate the trends obtained by averaging over 10 independent simulations of each model and the shaded regions represent the minimum and maximum value found at each $z$. The extent of these
shaded regions indicates the large variety of individual evolutionary histories that we find, particularly in models R300 and R1000. In Figures \ref{fig:10bhR300}, \ref{fig:10bhR1000}, \ref{fig:10bhSCA300}, and \ref{fig:10bhSCA1000}, we show the results obtained in each run of R300, R1000, SCA300, and SCA1000 models.

In model R300 (Figure \ref{fig:10bhR300}) a similar evolution is found for the total BH mass (black lines) among all the 10 runs, with the exception of runs (2) and (3) where the final BH mass does not exceed $2\times 10^8 \rm M_\odot$, as a consequence of the lower black hole accretion rate.

The relative contribution of medium-weight and light BH seed progenitors can be very different, reflecting the impact of the cosmological evolution of
the host galaxies on their birth conditions. The two populations provide almost equal contributions in some simulations, e.g. (9),  or largely different ones in other, such as run (7), where only a few light BH seed progenitors form.

A completely different scenario is found in model R1000, where in all the runs the final SMBH is $< 10^8 M_\odot$ at $z = 6.4$. This is a consequence of the smaller number (or absence, such as in run 3) of heavy BH seed progenitors formed and of their lower accretion rates. Indeed, a larger $J_{\rm cr}$ leads to milder radiative feedback and to more efficient (Pop III) star formation, increasing the number of light BH seed progenitors and decreasing the amount of gas that feeds BH growth. In 5 out of 10 simulations, the total mass contributed light BH seeds is comparable or exceeds the one provided by heavy or medium-weight progenitors. 

The 10 runs of the SCA300 (Figure \ref{fig:10bhSCA300}) and SCA1000 (Figure \ref{fig:10bhSCA1000}) models show similar evolutionary histories, with significantly larger numbers of medium-weight and heavy seed progenitors, with respect to R300 and R1000. This reflects the looser constraints set by SCA models on the formation of SMSs. A SMBH $>5\times 10^8 \rm M_\odot$ is predicted at $z=6.4$ in all cases and (almost) independently of the adopted $J_{\rm cr}$ threshold.

\bsp	
\label{lastpage}
\end{document}